\documentclass[11pt]{article}

\usepackage{fullpage}
\usepackage[utf8]{inputenc}
\usepackage{amsmath,amsfonts,amsthm,amssymb}
\usepackage{setspace}
\usepackage{fancyhdr}
\usepackage{lastpage}
\usepackage{extramarks}
\usepackage{chngpage}
\usepackage{soul,color}
\usepackage{graphicx,float,wrapfig}
\usepackage{listings}
\usepackage{xcolor}      
\usepackage{enumerate}
\usepackage{enumitem}
\usepackage{xspace}
\usepackage[T1]{fontenc}
\usepackage{subfiles}
\usepackage{bm}
\usepackage{tabularx}
\usepackage{multirow}

\usepackage{xcolor}
\usepackage{nameref}
\definecolor{ForestGreen}{rgb}{0.1333,0.5451,0.1333}
\definecolor{DarkRed}{rgb}{0.65,0,0}
\definecolor{Red}{rgb}{1,0,0}
\usepackage[linktocpage=true,
pagebackref=true,colorlinks,
linkcolor=DarkRed,citecolor=ForestGreen,
bookmarks,bookmarksopen,bookmarksnumbered]
{hyperref}
\usepackage{cleveref}
\usepackage{algpseudocode}
\usepackage{algorithm}  
\usepackage{algorithmicx} 
\usepackage{verbatim}
\usepackage{todonotes}
\usepackage{appendix}

\makeatletter
\def\namedlabel#1#2{\begingroup
    #2%
    \def\@currentlabel{#2}%
    \phantomsection\label{#1}\endgroup
}
\makeatother

\usepackage{thmtools}
\usepackage{thm-restate}

\declaretheorem[numberwithin=section]{theorem}
\declaretheorem[numberlike=theorem]{lemma}

\declaretheorem[numberlike=theorem]{corollary}

\declaretheorem[numberlike=theorem]{invariant}

\usepackage{mathrsfs}

\newcommand{\poly}{\operatorname{\text{{\rm poly}}}}
\newcommand{\polylog}{\operatorname{\text{{\rm polylog}}}}

\newcommand{\PRIMALELL}{\ensuremath{\textsc{Primal}^{(L)}}}
\newcommand{\PRIMAL}{\ensuremath{\textsc{Primal}}}
\newcommand{\DUALTWO}{\ensuremath{\textsc{Dual}^{(2)}}}
\newcommand{\maximalsets}{maximal edge-disjoint eligible open triangles}
\newcommand{\opentrianglequ}{partial triangle inequality}
\newcommand{\OPT}{\mathrm{OPT}}

\renewcommand{\cong}{\mathit{cong}}

\newcommand{\PTwo}{\mathcal{P}_2}
\newcommand{\half}{\frac{1}{2}}

\newcommand{\pivot}{\textsc{Pivot}}

\newcommand{\seqrounding}{\textsc{SeqRounding}}

\DeclareMathOperator*{\argmin}{arg\,min}

\Crefname{ALC@unique}{Line}{Lines}

\title{Breaking 3-Factor Approximation for Correlation Clustering in Polylogarithmic Rounds\thanks{This work was supported by NSF CCF-2008422.}} 
 \author{Nairen Cao\\
  Boston College\and 
 Shang-En Huang\\
 Boston College\and
 Hsin-Hao Su\\
 Boston College
}
\date{}

\DeclareMathSizes{10.95}{11}{8.5}{8}

\begin{document}

\maketitle

\begin{abstract}
In this paper, we study parallel algorithms for the correlation clustering problem, where every pair of two different entities is labeled with similar or dissimilar. The goal is to partition the entities into clusters to minimize the number of disagreements with the labels. Currently, all efficient parallel algorithms have an approximation ratio of at least 3. In comparison with the $1.994+\epsilon$ ratio achieved by polynomial-time sequential algorithms \cite{CLN22}, a significant gap exists.

We propose the first poly-logarithmic depth parallel algorithm that achieves a better approximation ratio than 3. Specifically, our algorithm computes a $(2.4+\epsilon)$-approximate solution and uses $\tilde{O}(m^{1.5})$ work. Additionally, it can be translated into a $\tilde{O}(m^{1.5})$-time sequential algorithm and a poly-logarithmic rounds sublinear-memory MPC algorithm with $\tilde{O}(m^{1.5})$ total memory.

Our approach is inspired by Awerbuch, Khandekar, and Rao's \cite{AKR12} length-constrained multi-commodity flow algorithm, where we develop an efficient parallel algorithm to solve a truncated correlation clustering linear program of Charikar, Guruswami, and Wirth \cite{CGW05}. Then we show the solution of the truncated linear program can be rounded with a factor of at most 2.4 loss by using the framework of \cite{chawla2015near}. Such a rounding framework can then be implemented using parallel pivot-based approaches (e.g. \cite{blelloch2012greedy,FischerN20}).

\end{abstract}
\thispagestyle{empty} 
\clearpage
\setcounter{page}{1}
\sloppy
\section{Introduction}
We study parallel algorithms for the correlation clustering problem introduced by Bansal, Blum, and Chawla~\cite{BBC04}, where the goal is to group similar entities and keep different entities apart. In the problem, we are given a complete graph $G = (V,E^{+} \cup E^{-})$ where $E^{+}$ are the edges labeled with $+$ (similar) and $E^{-}$ are the ones labeled with $-$ (different).   Given a clustering of the vertices, we say an edge $uv$ {\it disagrees} with the clustering if either (1) $uv$ is labeled with $-$ and $u,v$ are in the same cluster, or (2)  $uv$ is labeled with + and $u,v$ are in the different clusters.  The objective is to find a clustering that minimizes the number of edges that disagree. 

 In contrast to most other clustering methods, correlation clustering does not require the user to specify the number of clusters as the input. Due to its simplicity, this clustering method has various applications in spam detection, gene clustering, chat disentanglement, and co-reference resolution~\cite{CDK14, BGL14, ES09, ARS09, EIV07}. 
 

In the sequential setting, several algorithms have been developed with increasingly better approximation ratios \cite{BBC04, CGW05, ACN08,chawla2015near,CLN22}. In particular, \cite{ACN08} introduced the classic \pivot{} algorithm, which achieves an approximation factor of $3$. They also showed how to use it to round a linear programming (LP) solution with a 2.5-approximation ratio. Later, \cite{chawla2015near} improved the approximation factor to $2.06$. Recently, \cite{CLN22} designed a rounding algorithm based on the Sherali-Adams relaxation which achieves an approximation factor of $1.994+\epsilon$. The correlation clustering problem is also known to be APX-hard \cite{CGW05}. 

The exponential growth of data and advances in parallel architectures have motivated a long line of study on the parallel algorithms for the correlation clustering problem~\cite{blelloch2012greedy,CDK14,FischerN20,CCMU21,CLMNPT21,ACGMW21,AW22,CPU22,BCMT22}. Algorithms of particular interest are those with small, say, poly-logarithmic depths\footnote{The {\it work} of a parallel algorithm is the total number of primitive operations, and its {\it span} or {\it depth} is the length of the longest chain of sequential dependencies or, equivalently, the limit of parallel time as processors approach infinity.}, as having a small depth is usually the prerequisite of efficient implementations in modern parallel architectures. We summarize the results that fit into this category in \Cref{table:parallel_algs}. Many of the algorithms are designed for specific models such as the Massively Parallel Computation (MPC) model, the streaming model, and the local computation model (see \cite{BCMT22} for a comprehensive survey), but they can be easily transformed into algorithms with small depths (some with additional logarithmic factors). Noticeably, there has been a successful line of work based on parallelization of the \pivot{} algorithm of \cite{ACN08}, to which they culminated in constant rounds MPC algorithms \cite{BCMT22, CPU22}.

\begin{table}[h]\centering
\begin{tabular}{|l|l|l|l|}
\hline
Reference                 & Approx. Ratio & Rounds                & Method                 \\ \hline
\cite{AW22}               & $\approx $ 100000         & $1$              &  \multirow{2}{*}{sparse-dense decomposition}                     \\ \cline{1-3}
\cite{CLMNPT21}           & 701           & $O(1)$           &                             \\ \hline
\cite{ACGMW21}            & 3             & $O(\log \log n)$    & \multirow{7}{*}{\pivot-based} \\ \cline{1-3}
\cite{CCMU21}             & 3             & $O(\log \Delta \cdot \log \log n)$  &   
                 \\ \cline{1-3}
\cite{FischerN20}         & 3             & $O(\log n)$         &        
                  \\ \cline{1-3}
\cite{blelloch2012greedy} & 3             & $O(\log^2 n)$          &    
                  \\ \cline{1-3}
\cite{CDK14}              & $3+\epsilon$  & $O((\log n \cdot \log \Delta) / \epsilon)$ & 
                  \\ \cline{1-3}
\cite{CPU22}              & $3+\epsilon$ & $O(1)$  &
                  \\ \cline{1-3}
\cite{BCMT22}             & $3+\epsilon$  & $O(1/\epsilon)$     &
                  \\ \hline
\textbf{This paper.} & $2.4+\epsilon$ & $\poly(1/\epsilon,\log n)$ & LP + \pivot-based
                   \\ \hline
\end{tabular}\caption{\small Low-depth parallel algorithms for correlation clustering. The running times are stated with respect to the model of focus in each work.}\label{table:parallel_algs} 
\end{table}

However, these algorithms hit a barrier at the approximation factor of 3. A natural question is whether parallel algorithms of poly-logarithmic depths for achieving an approximation factor better than 3 exist. Indeed, \cite{BCMT22} also mentioned in the conclusion and open problem section, \begin{quote} ``It would be extremely interesting to study whether a low-round algorithm exists for solving the natural correlation clustering LP.'' -- as it would lead to algorithms with better than 3 approximation ratios. \end{quote} 

In this paper, we give the first a poly-logarithmic depth parallel algorithm that surpasses the 3-factor approximation. In previous literature, it is typical to use $m = |E^{+}|$ to denote the number of positive edges and to obtain bounds in terms of $m$, as it has been pointed out by \cite{CDK14} that it is common to have a much smaller number of positive edges than negative edges in practical applications.

\begin{theorem}
\label{theorem:maintheorem}
There exists a $\tilde{O}(\epsilon^{-4})$-depth parallel algorithm that achieves a $(2.4+\epsilon)$-approximation for the correlation clustering problem using a total work of $\tilde{O}(\epsilon^{-7}m^{1.5})$. \footnote{The bounds are stated for the PRAM CRCW model, although other variants induce at most a logarithmic factor.}\end{theorem}

Moreover, our parallel algorithms can also be simulated in the sublinear-space MPC model with a total space of $\tilde{O}(\epsilon^{-3}m^{1.5})$. We summarize this result in~\Cref{cor:mpc} and prove it in~\Cref{sec:mpc}.

\begin{corollary}\label{cor:mpc} There exists a $\tilde{O}(\epsilon^{-4})$-round sublinear memory MPC algorithm that computes a $(2.4+\epsilon)$-approximate solution for the correlation clustering problem using a total memory of $\tilde{O}(\epsilon^{-3}m^{1.5})$. 
\end{corollary}

In the sequential setting, all previous algorithms with approximation factors better than 3 would require solving the standard linear program by \cite{ACN08}, which would take at least $\Omega(n^5)$ time by the fastest solver today. As sequential algorithms can be directly obtained by simulating parallel algorithms, we also obtain a sequential algorithm whose running time equals to the work of our parallel algorithm:

\begin{corollary}\label{cor:seq} There exists a $\tilde{O}(\epsilon^{-7}m^{1.5})$ time sequential algorithm that computes a $(2.4+\epsilon)$-approximate solution for the correlation clustering problem.
\end{corollary}

Interestingly, the work bottleneck in \Cref{theorem:maintheorem} comes from the problem of finding a maximal set of edge-disjoint open triangles, where an open triangle is a triangle with 2 positive edges and 1 negative edge. We show that any combinatorial algorithm with better work than $O(m^{1.5-\epsilon})$ for any constant $\epsilon > 0$ for the problem would refute a conjecture on the Boolean matrix multiplication problem \cite{AW14,williams2010subcubic}. 

\paragraph{Further Related Works on Correlation Clustering.}
There have also been studies of the correlation clustering problem on non-complete graphs where there may be missing labels among pairs of vertices. In such a setting, \cite{DEFI06} gave an $O(\log n)$-approximation sequential algorithm for the problem. For the agreement maximization version of the problem where the goal is to maximize the number of edges in agreement with the partition, \cite{Swamy04} and \cite{CGW05} gave a 0.7664-approximation algorithm and a 0.7666-approximation algorithm for the problem respectively. The correlation clustering problem has also been studied in online settings \cite{MSS10, LMVWZ21,CLMP22}, and settings with differential privacy guarantees \cite{BEK21,CCLMNPT22,Liu22} and local guarantees \cite{CGS17,PM18, KMZ19,JKMM21}. 

\subsection{Technical Challenges}
As mentioned in \cite{BCMT22}, the bottleneck for achieving efficient parallel algorithms with an approximation factor better than 3 is in solving the standard correlation clustering LP (P1). The existing sequential 2.06-approximation algorithm by \cite{chawla2015near} and sequential 2.5-approximation algorithm by \cite{ACN08} both start with a fractional solution to (P1). Note that the 2.06 approximation is nearly optimal as the integrality to the LP is 2. \cite{BCMT22} pointed out that once a fractional solution of the linear program is obtained, one can create an instance in the spirit of the 2.06-approximate algorithm of \cite{chawla2015near} such that running the $\pivot$-style parallel algorithms on the instance yields the same approximation factor. 

\paragraph{Standard Correlation Clustering LP}
\begin{equation}
\begin{aligned}
    \text{minimize }\ \  & \sum_{(u, v)\in E^+} x_{uv} + \sum_{(u, v)\in E^-} (1-x_{uv}) \\
    \text{subject to }\ \  & \begin{aligned}[t] 
    & x_{uw} + x_{wv} \ge x_{uv} && \ \ \forall u, v, w\in V\\
    & x_{uu} = 0 && \ \ \forall u\in V\\
    & x_{uv} \in [0, 1] && \ \ \forall (u, v)\in E^+\cup E^-
    \end{aligned} 
\end{aligned}\tag{P1}
\end{equation}

The most efficient solver by using the interior point method today \cite{NM94,LeeS14} would require at least $\Omega(\sqrt{N})$ iterations, where $N$ is the number of variables, which in our case is $\Theta(n^2)$. This is even not counting the fact each iteration involves complicated volume computation or linear system solving. In the sequential setting, the best known algorithm for solving a general linear program runs in $\tilde{O}(N^{\omega} + N^{2.5-\alpha/2} + N^{13/6})$ time by \cite{CLS19}, where $\omega \sim 2.37$ and $\alpha \sim 0.31$ are the current best-known exponent and dual exponent of matrix multiplication.

\paragraph{The Cut-Flow View.} 
As solving the standard linear program of correlation clustering using the current tools requires at least polynomial iterations, we take a detour from the problem.
The correlation clustering problem is known to have a strong connection with the multicut problem \cite{DEFI06}. 
Charikar, Guruswami, and Wirth \cite{CGW05} gave the following alternative linear program formulation which captures such a connection. \\[8pt]
\begin{tabular}{rll|rll}
\multicolumn{3}{c|}{(\PRIMAL)} & \multicolumn{3}{c}{(\textsc{Dual})}\\[6pt]
    min & \multicolumn{2}{l|}{$\sum_{(u, v)\in E^+ \cup E^-} z_{uv}$} & 
    max & \multicolumn{2}{l}{$\sum_{P\in \mathcal{P}} y_P$}  \\[6pt]
    s.t. & $z_{uv} + \sum_{e\in P} z_e \ge 1$  & $\forall uv\in E^- \ \forall P\in \mathcal{P}_{uv}$ & 
    s.t. & $\sum_{P\ni e} y_P\le 1$ & $\forall e\in E^+$ \\[6pt]
    &   & &
    & $\sum_{P\in\mathcal{P}_{uv}} y_P \le 1$ & $\forall uv\in E^-$ \\[6pt]
    & $z_{uv}\ge 0$  & $\forall uv\in E^+ \cup E^-$ & 
    & $y_P\ge 0$ & $\forall P\in \mathcal{P}$ 
\end{tabular}
\medskip \\
where $\mathcal{P}_{uv}$ be the collection of paths in $G^+$ that connects $u$ and $v$ and $\mathcal{P}=\cup_{uv\in E^-} \mathcal{P}_{uv}$.

\cite{CGW05} showed that an optimal solution to (\PRIMAL) is an optimal solution to the standard linear program (P1). In this cut-flow view, the dual linear program (\textsc{Dual}) can be seen as a variant of the multi-commodity flow problem on $E^{+}$ where each negative edge $(u,v) \in E^{-}$ is a source-sink pair. The objective becomes routing as much flow as possible between the source-sink pairs under the constraints that (1) each edge has capacity 1, and (2) at most 1 unit flow can be routed between each source-sink pair.    Although the multi-commodity flow problem has been studied extensively, unfortunately, there are no known parallel algorithms that can solve it in poly-logarithmic iterations.


\subsection{Our Approach}
We truncate the Cut-Flow View linear program, by only keeping the path constraints for bounded-hop paths. The following is the truncated primal program. 

\paragraph{The $L$-hop Cut-Flow View.} Fix a positive integer $L$. For any negative edge $(u, v)\in E^-$, we denote $\mathcal{P}_{uv}^{(L)}$ to be the paths in $\mathcal{P}_{uv}$ with at most $L$ edges (hops). Similarly we define $\mathcal{P}_L = \cup_{uv\in E^-} \mathcal{P}_{uv}^{(L)}$.~\\[8pt]
\begin{tabular}{rll|rll}
\multicolumn{3}{c|}{(\PRIMALELL)} & \multicolumn{3}{c}{($\textsc{Dual}^{(L)}$)} \\[6pt]
min & \multicolumn{2}{l|}{$\sum_{(u, v)\in E^+ \cup E^-} z_{uv}$} & 
max & \multicolumn{2}{l}{$\sum_{P\in \mathcal{P}_L} y_P$} \\[6pt]
s.t. & $z_{uv} + \sum_{e\in P_{uv}} z_e \ge 1$ & $\forall uv\in E^-\ \forall P\in \mathcal{P}_{uv}^{(L)}$ &
s.t. & $\sum_{P\ni e} y_P \le 1$ & $\forall e\in E^+$ \\[6pt]
    & & &
    & $\sum_{P\in \mathcal{P}_{uv}^{(L)}} y_P \le 1$ & $\forall uv\in E^-$ \\[6pt]
    & $z_{uv}\ge 0$ &  $\forall uv\in E^+ \cup E^-$ &
    & $y_P \ge 0$ & $\forall P\in \mathcal{P}_L$
\end{tabular}

\medskip

The corresponding dual program is also a variant of the multi-commodity flow problem where we can only route flow along paths of at most $L$-hops between the source sink pairs. The hop-constrained multi-commodity flow problem has been studied by Awerbuch, Khandekar, and Rao \cite{AKR12}, where they developed an algorithm that runs in $\tilde{O}(L)$ iterations (note that the standard multiplicative weight update method will take too much time because the width can be very large). Although our dual program has an extra constraint, their result suggests the possibility of having a poly-logarithmic round parallel algorithm, for $L$ up to $\poly(\log n)$. However, even if it works,  such an approach still faces the following challenges:

\begin{enumerate}[leftmargin = *]
\item \label{clng1} First, it is unclear how far the optimal solution of \PRIMALELL{} is from that of \PRIMAL. Moreover, we will need to develop a procedure to convert a solution of \PRIMALELL{} to, say some approximation solution of \PRIMAL.

\item \label{clng2} Second, all the existing rounding algorithms are based on the standard correlation clustering LP. Even if we can convert an optimal solution of \PRIMALELL{} to an approximate solution of \PRIMAL, it is unclear if it will satisfy the constraints of the standard correlation clustering LP. Note that \cite{CGW05} only showed an optimal solution will satisfy the constraints, but an approximate solution will not necessarily do so.

\item \label{clng3} Third, as mentioned previously, we hope to obtain an upper bound on the total work in terms of $m = |E^{+}|$, as it will result in a faster sequential algorithm on graphs with sparse positive edges. The challenge is two-fold. First, in the fractional primal solution, there can be much more than $m$ negative edges with non-zero values. This means the input to the rounding algorithm could have $\Omega(n^2)$ edges in  the worst case.  Second, in the algorithm of \cite{AKR12}, each iteration requires computing an approximate blocking flow for every source-sink pair. Potentially, there can be $\Omega(n^2)$ such source-sink pairs. For each source-sink pair, the number of paths of length at most $L$ can be as large as $\Omega(n^{L-1})$ between each pair. If we assign a processor to each of the paths directly, the work would be at least $\Omega(n^{L+1})$.

\end{enumerate}

Surprisingly, we found that it is possible to overcome the first two challenges altogether by directly rounding a solution of \PRIMALELL{} for $L=2$ to an integral correlation clustering solution. Our new rounding procedure obtains an approximation factor of 2.4. 

\paragraph{Two-Hop Primal ($L=2$).}
\begin{equation}
\begin{aligned}
    \text{minimize }\ \  & \sum_{(u, v)\in E^+ \cup E^-} z_{uv} \\
    \text{subject to }\ \  & \begin{aligned}[t]
    & z_{uv} + z_{uw} + z_{wv} \ge 1 && \ \ 
    \forall (u, v)\in E^-\text{ and } (u, w), (w, v)\in E^+ \\
    & z_{uv} \ge 0 && \ \ \forall (u, v)\in E^+\cup E^-
    \end{aligned} 
\end{aligned}\tag{$\PRIMAL^{(2)}$}
\end{equation}

\paragraph{Two-Hop Dual.} Let $\PTwo$ be the collection of length-two paths $(u, w, v)$ such that $(u, w), (w, v)\in E^+$ and $(u, v)\in E^-$. 
\begin{equation}
\begin{aligned}
    \text{maximize }\ \  & \sum_{(u, w, v)\in \PTwo} y_{(u, w, v)} \\
    \text{subject to }\ \  & \begin{aligned}[t]
    & \sum_{w: (u, w, v)\in \PTwo} y_{(u, w, v)} \le 1 && \ \ 
    \forall (u, v)\in E^- \\
    & \sum_{v': (u, w, v')\in \PTwo} y_{(u, w, v')} + \sum_{u': (u', w, u)\in\PTwo} y_{(u', w, u)}  \le 1 && \ \ \forall (u, w)\in E^+\\
    & y_{(u, w, v)}\ge 0 && \ \ \forall (u, w, v)\in \PTwo
    \end{aligned} 
\end{aligned}\tag{$\DUALTWO$}
\end{equation}

\begin{lemma}\label{lem:rounding} There exists a $\poly(\log n)$-time parallel algorithm that converts a fractional solution of $\PRIMAL^{(2)}$ to a clustering where the number of disagreements is at most 2.4 times the object value of the fractional solution in expectation. The total work of the algorithm is quasi-linear in the number of non-zero terms of the fractional solution.\end{lemma}

\begin{corollary}Let $OPT(\cdot)$ denote the optimal value of a linear program. Let $\PRIMAL_{I}$ be the integral version of $\PRIMAL$. We have $$OPT(\PRIMAL) \leq OPT(\PRIMAL_{I}) \leq 2.4 \cdot OPT(\PRIMAL^{(2)})$$\end{corollary}

An {\it open triangle} is $(x,y,z)$, where $(x,y),(y,z) \in E^{+}$ and $(x,z) \in E^{-}$. A {\it closed triangle} is $(x,y,z)$, where $(x,y),(y,z), (x,z) \in E^{+}$. We note that $\PTwo$ can also be seen as a collection of open triangles. The above dual program can be viewed as an open triangle packing problem, where the goal is to pack as many open triangles as possible fractionally, subject to the condition that for each $e \in E^{+} \cup E^{-}$, the sum of the values over all the triangles containing $e$ is at most 1. 

We note that recently, Veldt \cite{Veldt22}, also considered the same primal LP for correlation clustering. He developed a 4-approximation rounding algorithm for correlation clustering from the LP. Moreover, he noted that such a linear program can be solved faster {\it empirically} by LP solvers. Although we use the same LP, the paths of how we arrive at such an LP are different, yet coincidental. Veldt \cite{Veldt22} drew and strengthened an interesting connection between correlation clustering and {\it strong triadic closure labeling} \cite{ST14, GM20}, while we derive such a linear program from the lens of efficient optimization algorithms. We also emphasize that the LP is different than the standard LP (P1) studied by \cite{ACN08} as they consider {\it all} $(u,w,v)$ tuples in the constraint instead of only the ones in $\mathcal{P}_2$.

\paragraph{A $2.4$-Approximation Rounding Algorithm.} Ailon, Charikar, and Newman~\cite{ACN08} first proposed a 2.5-approximation algorithm by rounding the linear programming solution $\{ x_e \}_{e\in E^{+} \cup E^{-}}$ of (P1). They interpret $x_e$ as the probability that the edge $e$ should be cut, with higher values of $x_e$ indicating a lower probability that the endpoints $e$ should be included in the same cluster. Their algorithm is based on their \pivot{} algorithm and works as follows: Choose a random vertex $w$ as the pivot. Put every other node $v$ into the cluster of $w$ with probability $1-x_{wv}$. Remove the cluster and then repeat on the remaining graph.

Later, Chawla et al.~\cite{chawla2015near} improved this method to a $2.06$ approximation ratio by making the crucial observation that clustering a node $u$ with probability $1- x_{uw}$ may not lead to the best approximation ratio. Instead, they cluster the node with probability $1 - f(x_{uw})$, for some carefully chosen function $f$. 

The key in upper bounding the approximation ratio the \pivot-based rounding algorithm, as developed in \cite{ACN08, chawla2015near} and elaborated in~\cite{CLN22}, is in bounding the following ratio: 
$$
\rho(u,v,w) = \frac{cost(u, v \mid w) + cost(u, w \mid v) + cost(w, v \mid u)}{lp(u, v \mid w) + lp(u, w \mid v) + lp(w, v \mid u)}
$$
for every triangle $(u, v, w)$. The term $cost(u, v \mid w)$ denotes the probability that the disagreement of $(u, v)$ occurs when $w$ is chosen as the pivot, whereas $lp(u, v \mid w)$ denotes the probability that the edge $uv$ is removed from the graph when $w$ is chosen as the pivot, multiplied by the contribution of the edge $uv$ in the LP. Intuitively, the numerator can be understood as the actual cost caused by the algorithm, while as the denominator can be understood as the cost in the LP that we are charging to. Their analysis shows that $\max_{(u,v,w)}\rho(u,v,w)$ is an upper bound of the approximation ratio of the rounding algorithm.



\cite{chawla2015near}  noted that $\rho(u,v,w)$ is a multivariate polynomial over $x_{uv}, x_{uw}, x_{vw}$ and conducted a case-by-case analysis of $\rho(\cdot)$ over four types of triangles based on the sign of their edges: $(-,-,-), (+,-,-), (+,+,-), (+,+,+)$. Their carefully designed rounding function, $f(\cdot)$, led to an upper bound of $\rho(u,v,w)$ by $2.06$ for every type of triangle. The constraints of (P1), namely the triangle inequality $x_{uw} + x_{vw} \geq x_{uv}$ over all the triples $(u,w,v)$, play a critical role in their analysis.



One of our main technical contributions is that we show that we do not need triangle inequality for all types of triangles to obtain a good approximate ratio. Specifically, we show that for all triangles $(u, w, v)$ such that $uw, vw \in E^{+}$ and $uv \in E^{-}$ (the $(+,+,-)$ triangle), if we have $x_{uw} + x_{vw} \geq x_{uv}$, then it is possible to obtain an upper bound of $2.4$ on $\rho(u,v,w)$ for {\it all} types of triangles, by using a different rounding function $f$.  
In addition, we show that $2.4$ is the best ratio one can obtain when under such a framework.

Focusing solely on $(+,+, -)$ triangles greatly simplifies the process of solving linear programs.
Instead of solving (P1), the solution to our $\PRIMAL^{(2)}$ provides us with the triangle inequality for $(+, +, -)$ triangles. This can be done by setting $x_e = z_{e}$ for all positive edges and $x_{e} = 1 - z_{e}$ for all negative edges. Thus, a feasible solution from $\PRIMAL^{(2)}$ can be converted into an assignment that satisfies the triangle inequality for $(+, +, -)$ triangles. 





\paragraph{Solving $\PRIMAL^{(2)}$.}
Awerbuch, Khandekar, and Rao~\cite{AKR12} proposed a distributed \emph{approximate steepest descendant framework} that gives approximate solutions to multi-commodity flow problems efficiently.
In their framework, there is a \emph{convex} length function with respect to the congestion $\cong_e$ of each edge (say $(m^{1/\epsilon})^{\cong_e}$),
and the objective is to minimize the sum $\Phi$ of all edge lengths while maximizing the total flow $\sum y_P$. Let $k$ be the number of commodities.
In each step, for each commodity, the algorithm chooses a set of \emph{approximately shortest} source-sink paths and runs a \emph{blocking flow} through these paths, where the capacities of blocking flow are set to be tiny, roughly $\epsilon / k$.
By the pigeonhole principle, all approximately shortest source-sink paths will be eliminated after a certain number of steps.
It turns out that, by choosing only the shortest paths in each step, one can bound the growth rate of $\Phi$. Furthermore, these edge lengths, divided by the shortest paths' length, can be used to define a feasible primal solution.
This establishes a $(1+O(\epsilon))$ factor difference between primal and dual objective values,
which certifies a desired $(1+O(\epsilon))$ approximation ratio.

Inspired by the steepest descendant framework,
we extend the length function to not only the positive edges (the edges presented in the input) but also the negative edges.
Moreover, each source-sink pair corresponds to a negative edge $uv\in E^-$ where there exist two-hop paths in $E^+$ connecting $u$ and $v$, forming $(+, +, -)$ triangles. Sending a flow from $u$ to $v$ is then equivalent to adding a circulation to a $(+, +, -)$ triangle involving $uv$.
This makes the entire framework applicable to solve $\PRIMAL^{(2)}$ and $\DUALTWO$.

 However, the third challenge still remains, even if we restrict the number of hops to $L=2$.
 This is mainly because there can be $k = \Omega(n^2)$ source-sink pairs, one per negative edge.
In the algorithm of \cite{AKR12}, they compute an approximate shortest blocking flow for every source-sink pair. This can cause the algorithm to send flow through $\Omega(n^2)$ edges in a single iteration. Additionally, computing the blocking flow for $k = \Omega(n^2)$ may require a significant amount of work.
If one would like to efficiently utilize \Cref{lem:rounding}, which requires quasi-linear work in the number of non-zero terms of the fractional solution, we have to first ensure that our fractional solution has a $\tilde{O}(m)$-sized support.

To resolve this, our idea is to mix up all the commodities. That is, we let each commodity block other commodities, instead of computing the blocking flow per commodity. In the view of triangles, this corresponds to computing a maximal set of edge-disjoint open triangles. There can be at most $O(m)$ edge-disjoint open triangles in such a set because each open triangle consumes two positive edges. At the same time, instead of sending $\epsilon / k$ flow for each commodity, we can send up to $\epsilon / \log m$ flow for each commodity in the maximal set. This, combined with the fact each triangle contains at least two positive edges and one negative edge, gives an upper bound of $\tilde{O}(m)$ on the number of non-zero duals. However, these do not necessarily translate directly to bounds on the number of non-zero or relatively small primal variables. We devise an additional simple post-processing step to construct a primal solution where all but $\tilde{O}(m)$ primal variables are small enough to be truncated.


\begin{restatable}{lemma}{lpmainlemma}
\label{lemma:lpmainlemma}
Given the set $E^+$ of $m$ positive edges and a parameter $\epsilon > 0$, there exists a parallel algorithm that computes an $(1+\epsilon)$-approximate solution to $\PRIMAL^{(2)}$ using $\tilde{O}(\epsilon^{-7} m^{1.5})$ work and $\tilde{O}(\epsilon^{-4})$ span. In addition, the support size of the solution is  at most $\tilde{O}(\epsilon^{-2} m)$.
\end{restatable}

The last mile to our $\tilde{O}(m^{1.5})$-work $\poly(\log n)$-depth parallel algorithm is the problem of computing a maximal set of edge-disjoint open triangles. Although, it can be shown that the number of closed triangles is upper bounded by $\tilde{O}(m^{1.5})$, it is not necessarily the case for open triangles. The number of open triangles can be as large as $\Omega(n^2)$ even when there are $O(n)$ edges (e.g.~a star of positive edges).
Inspired by the triangle enumeration algorithms,
we observe that it is possible to compute a maximal set of edge-disjoint open triangles without checking all open triangles. To this end, we obtain a parallel algorithm that gradually searches for more edge-disjoint open triangles in rounds.
In each round, the algorithm explores a collection of open triangles $\mathcal{C}$ and runs a maximal independent set (MIS) algorithm in the conflict graph built on $\mathcal{C}$.
As long as these open triangles in $\mathcal{C}$ are edge-disjoint to the current found set $S$ of edge-disjoint open triangles,
the open triangles selected in the MIS can be added to $S$ and all open triangles in $S$ are still edge-disjoint.
By carefully controlling the exploration rate, we obtain the parallel algorithm with desired work and depth, summarized below in \Cref{lemma:intro-max-triangle} and proved in \Cref{section:maximal-edge-disjoint-triangles}.

\begin{restatable}{lemma}{maxtriangle}
\label{lemma:intro-max-triangle}
Let  $G=(V, E^+\cup E_{>1}^-)$, and $l:\binom{V}{2} \to [1, \infty)$ be a length function with $l(u, v) = 1$ for $uv \in \binom{V}{2} \setminus (E^{+} \cup E_{>1}^{-1})$, and $L > 0$ be a length limit. Let $m_f = |E^+\cup E_{>1}^-|$. Then, there exists a parallel algorithm such that, in $O(m_f^{1.5}\log^3 n)$  work and $O(\log^3 n)$ span, the algorithm returns a maximal edge-disjoint set $S$ of open triangles with length less than $L$.
\end{restatable}


\paragraph{Conditional Lower Bound for Maximal Edge-Disjoint Open Triangles.}
Unfortunately, it seems that $\tilde{O}(m^{1.5})$ is the best total work one can hope for, if we want a combinatorial algorithm that computes a maximal set of edge-disjoint open triangles.
In terms of the number of total input edges $m$, the problem of finding a maximal set of edge-disjoint open triangles could be as hard as searching for just one (regular, non-open) triangle.
The latter problem (triangle detection) has a conditional lower bound based on the combinatorial Boolean matrix multiplication (BMM) problem~\cite{williams2010subcubic}.
In \Cref{subsection:hardness} we give a randomized reduction from the triangle detection problem to our maximal open triangle problem. Such a reduction also implies that {\it any} algorithms using $O(m^{1.185-\delta})$ work for any constant $\delta > 0$ for our problem would lead to an improvement over the best known algorithm for the triangle detection problem \cite{bjorklund2014listing}.




\subsection{Open Problems}
We give the first poly-logarithmic depth parallel algorithm that achieves an approximation ratio better than 3. We hope that our work can shed some light on the search of low-round algorithms with an approximation ratio of less than 3 in other models that exploit parallelism, such as the streaming model, the MPC model (with nearly-linear total memory), and the CONGEST model.
The main bottleneck in adapting our algorithm to those models is in finding a maximal set of edge-disjoint open triangles. 
It is interesting to investigate whether there are ways to sparsify the process in these models.  

We have shown that the optimal solution of $\PRIMAL^{(2)}$ is at most 2.4 times that of $\PRIMAL$. It is interesting to see if there are tighter relations between $\PRIMAL^{(L)}$ and $\PRIMAL$ for $L > 2$. Note that \cite{CGW05} showed the optimal values to both linear programs are the same when $L=n-1$.
Perhaps one may obtain a trade-off between $L$ and the quality of the solution.
If this is the case, we may be able to obtain a result that exhibits a trade-off between the running time/work of the algorithms and the quality of the solutions. 

Finally, we have shown a conditional lower bound on the problem of finding a maximal set of edge-disjoint open triangles. It is interesting to directly investigate the fine-grained complexity of the $c$-approximate correlation clustering for $c < 3$. We already know that the 3-approximation algorithm of \cite{ACN08} can be implemented in $\tilde{O}(m)$ time in the sequential setting, and $\tilde{O}(m)$ total work in the parallel setting~\cite{FischerN20}.
Now the question is whether it is possible to obtain a nearly-linear time algorithm for $c$-approximate algorithm for $c < 3$ or it can be shown to be (conditionally)-hard.

\section{Preliminaries}
Let $m = |E^{+}|$ be the number of positive edges in the unweighted and undirected 
input graph $G^{+} = (V, E^{+})$.
We will use the triple $(u, w, v)$ to denote the triangle with edges $(u, v)$, $(v, w)$ and $(u, w)$.
We note that the same triangle and the same undirected edge can be identified in different ways.
In particular, $(u, w, v)$ and $(v, w, u)$ refer to the same triangle, and $(u, v)$, $(v, u)$, $uv$, and $vu$ refer to the same edge.
Given an edge $e$ and a triangle $(u, w, v)$, we say that $e$ is \emph{on the triangle}, denoted as $e \in (u, w, v)$, if $e$ is one of the edges of $(u, v)$, $(v, w)$ and $(u, w)$.

Sometimes we denote the same triangle with an {\it ordered} 3-edge triple $(uw,wv,vu)$ when we are mapping certain attributes to the edges. Let $s_{uw}, s_{wv}, s_{vu} \in \{+,-\}$, a $(s_{uw}, s_{wv}, s_{vu})$ triangle is where $uw \in E^{s_{uw}}$, $wv \in E^{s_{wv}}$, and $vu \in E^{s_{vu}}$. For example, a $(+, +, -)$ triangle is a triangle $(uw, wv, vu)$ such that $uw, wv \in E^{+}$ and $vu \in E^{-}$. Such a triangle is called an \emph{open triangle}. Although $\PTwo$ was defined be the collection of length-two paths $(u, w, v)$ such that $(u, w), (w, v)\in E^+$ and $(u, v)\in E^-$, we can also view it as a collection of open triangles. A $(+,+,+)$ triangle is called a \emph{closed triangle}.



\paragraph{Assumptions.} We assume without loss of generality that the graph $G^{+} = (V, E^{+})$ is connected.
Otherwise, we can process each connected component induced by positive edges separately.
Also, we assume that $G^{+}$ is not a complete graph, so the optimal objective value is at least 1. Otherwise, we may just output the entire graph as a cluster.

\section{An $(1+\epsilon)$-Approximation Algorithm for $\PRIMAL^{(2)}$}
\label{sec:trianglepacking}
\label{sec:primalalgorithm}

In this section, we propose \Cref{alg:lp}, an algorithm
that computes a $(1 + \epsilon)$-approximate solution $\{ z_{uv} \}$ for $\PRIMAL^{(2)}$.
Our algorithm is inspired by the distributed steepest descent framework~\cite{AKR12} for the \emph{most beneficial flow} (MBF),
and an earlier sequential multicommodity flow algorithm by Garg and K\"onemann~\cite{GargK07}.
In our case, 
as mentioned in the introduction,
the algorithm focuses on triples in $\PTwo$ and sends flows along the \emph{most beneficial triangle}.

The algorithm runs in \emph{iterations}.
Intuitively, 
in each iteration $t$, 
the algorithm seeks a set $S$ of the approximately shortest length triangles from $\PTwo$.
Then, the algorithm pushes some tiny flow along each triangle, which by correspondence (see~\Cref{obser:algorithminvariant} below) increases the length of each edge with a multiplicative factor of $\exp(\epsilon)\approx 1+\epsilon$.
The iterations end once the total length of each edge exceeds a certain value, and the algorithm is then able to produce the $(1+O(\epsilon))$-approximate solutions to both $\PRIMAL^{(2)}$ and $\DUALTWO$.

\paragraph{Explicitly Maintained Variables and Invariants.}
Our algorithm mainly operates on $\DUALTWO$, that is, 
the algorithm explicitly stores all non-zero $y$ values for triangles in $\PTwo$. 
For the ease of analysis, we will use $y^{(t)}_{(u,w,v)}$
to denote the dual variables at the beginning of the iteration $t$.
If we treat each $y^{(t)}_{(u,w,v)}$ as a circulation on its own commodity, then it makes sense to define \emph{congestion} of an edge $e$, to be the sum of all flow values passing through that edge. Specifically, for each edge $e\in E^+\cup E^-$ we define $\cong^{(t)}_e = \sum_{(u,w,v)\in\PTwo, e \in (u, w, v)} y^{(t)}_{(u, w, v)}$.
The algorithm also explicitly maintains the \emph{length} of an edge $e$, which is defined by $l^{(t)}_{e} = (m^{1/\epsilon})^{\cong_{e}^{(t)}}$.
This leads to a definition for the length of a triangle $(u,w,v)$ in $t$-th iteration to be $l^{(t)}_{(u,w,v)} := l^{(t)}_{wu} + l^{(t)}_{wv} + l^{(t)}_{uv}$.
Moreover, the algorithm maintains a variable $\alpha^{(t)}$
for lower bounding the shortest open triangle.
Throughout the execution, the algorithm maintains the following invariant between the dual variables, length variables, and the shortest open triangle estimate:


\begin{invariant}
\label{obser:algorithminvariant}
At the beginning of any iteration $t$,
\begin{enumerate}[itemsep=-6pt]
    \item For all $(u, w, v) \in \PTwo$, $l^{(t)}_{(u, w, v)} \geq \alpha^{(t)}$,
    \item For any edge $e \in E^{+} \cup E^{-}$, $l^{(t)}_{e} = \big(m^{1/\epsilon}\big)^{\cong_e^{(t)}}$.
\end{enumerate}
\end{invariant}

\paragraph{Initialization.} 
Initially, the algorithm sets the dual variable $y^{(0)}_{(u, w, v)} \gets 0$ for each triangle $(u, w, v) \in \PTwo$.
Each edge has an edge length $l^{(0)}_e := (m^{1/\epsilon})^{\cong^{(0)}_e}$ which has an initial value $1$.
Since $\alpha^{(0)}$ is a lower bound for the shortest open triangle, we can safely set $\alpha^{(0)} \gets 3$ initially.

\paragraph{Termination Condition.}
Since the congestion of each edge is non-decreasing, the length of each edge is also non-decreasing.
The algorithm terminates when the total length of all edges becomes too large. Specifically, we define the following potential function
$$\Phi(t)=\sum_{e\in E^+\cup E^-}(m^{1/\epsilon})^{\cong_e^{(t)}}$$
and terminate the algorithm once $\Phi(t)$ surpasses $m^{1/\epsilon} / \exp(\epsilon)$.

\begin{algorithm}[htbp]
\caption{A $(1+O(\epsilon))$-approximate algorithm for $\PRIMAL^{(2)}$.\\
\textbf{Input}: A vertex set $V$, a set of $m$ undirected unweighted edges $E^+$,
and a parameter $\epsilon>0$.\\
\textbf{Output}: An $(1+O(\epsilon))$-approximate solution $\{z_{uv}\} \in [0,1]^{\binom{V}{2}}$ to $\PRIMAL^{(2)}$
\\
\textbf{Auxiliary Information}: $E^-:={V\choose 2}\setminus E^+$; $\PTwo :=$ the set of open triangles. 
}
\label{alg:lp}
\begin{algorithmic}[1]
\Function {ParallelSteepestDescent}{$G^+=(V, E^+), \epsilon$}
\Statex {\small \textit{\underline{$\triangleright$ Initialization}}}
    \State $t\gets 0$ and $\alpha^{(0)} \gets 3$ and $T_{\min} \gets +\infty$
    \State $l_{uv}^{(0)} \gets 1$ for all $(u, v)\in E^+\cup E^-$.
    \State $y_{(u,w,v)}^{(0)}\gets 0$ for all $(u,w,v)\in \PTwo$.
    \medskip
\Statex {\small \textit{\underline{$\triangleright$ Compute the primal and dual values}}}
    \While{$\Phi(t):=\sum_{uv} l_{uv}^{(t)} < m^{1/\epsilon}/\exp(\epsilon)$}\label{ln:1-compute-sum}
    \State If $\Phi(t) \geq m^3 / \epsilon $, set $T_{\min} \gets \min(T_{\min}, t)$
    \State Let $\PTwo' := \{(u,w,v)\in \PTwo \ |\ l_{(u,w,v)}^{(t)} < (1+\epsilon)\alpha^{(t)}\}$ be the set of eligible triangles;\label{ln:1-max-tri}
    \Statex \hspace{3em}Compute any maximal edge-disjoint set $S$ of $\PTwo'$. \Comment{See \Cref{section:maximal-edge-disjoint-triangles}.}
    \If{$S\neq \emptyset$}
    \State For all $(u, w, v)\in \PTwo$, set $y^{(t+1)}_{(u,w,v)}\gets \left\{\begin{array}{ll}   y^{(t)}_{(u,w,v)} + \epsilon^2/\ln m &  \text{ if $(u,w,v)\in S$,} \\
    y^{(t)}_{(u,w,v)} & \text{ otherwise.}
    \end{array}\right.$ \label{ln:1-sendflow}
    \State For all $(u, v)\in E$, set $l^{(t+1)}_{uv} \gets \left\{\begin{array}{ll}    l^{(t)}_{uv}\cdot \exp(\epsilon)
    &  \ \ \text{ if $(u, v)$ occur in some triangle in $S$,}\\
    l^{(t)}_{uv} & \ \ \text{ otherwise.}
    \end{array}\right.$
    \State $\alpha^{(t+1)}\gets \alpha^{(t)}$
    \Else 
    \State $\alpha^{(t+1)}\gets (1+\epsilon)\alpha^{(t)}$ \Comment{Update the lower bound estimate of the shortest triangle.}
    \EndIf
    \State $t\gets t+1$
    \EndWhile
    \medskip
    \Statex {\small \textit{\underline{$\triangleright$ Compute the primal solution}}}
    \State $T \gets \displaystyle\argmin_{ t \ge T_{\min}} \frac{\Phi(t)}{\alpha^{(t)}} $\label{ln:1-primal-sol}
    \State For $(u, v) \in E^{+} \cup E^{-}$, set $z_{uv}^{(T)} \gets \displaystyle\frac{l^{(T)}_{uv}}{\alpha^{(T)}}$ \label{alg:lpsetvalue}
    \State For $(u, v) \in E^{+}$, set $z_{uv} \leftarrow \min\left\{z^{(T)}_{uv} + \displaystyle\frac{\epsilon}{m}, 1\right\}$. \label{ln:finalplus}
    \State For $(u, v) \in E^{-}$, set $z_{uv} \gets \left\{\begin{array}{ll} 
    \min\left\{ z^{(T)}_{uv}, 1\right\} \label{ln:finalminus}
    &  \ \ \text{ if $l^{(T)}_{uv} > 1$,}\\
    0 & \ \ \text{ otherwise.}
    \end{array}\right.$\label{ln:trunc-ze}
    \State \Return $\{z_{uv}\}$
    \EndFunction
\end{algorithmic}
\end{algorithm}

\paragraph{Iterations.}
Within the $t$-th iteration of the algorithm, the algorithm attempts to send flows through the most beneficial triangle.
Specifically, we identify some triangle in $\PTwo$ with $(1+ \epsilon)$-approximate shortest distance and increase its dual value (and the corresponding edge lengths).
To accelerate the process, instead of sending flow through the most beneficial triangle one by one, the algorithm repeatedly selects a \emph{maximal} edge-disjoint set of triangles $S\subseteq \PTwo$ such that
 $l^{(t)}_{(u, w, v)} < (1+\epsilon)\alpha^{(t)}$ for every triangle $(u, w, v)\in S$, where $l^{(t)}$ is the current length function.
Then, the algorithm increases the dual variables $y^{(t)}_{(u, w, v)}$ for all triangles $(u, w, v)\in S$ by a fixed amount  $\Delta y^{(t)}_{(u,w,v)} := \epsilon^2/\ln m$.
To maintain \Cref{obser:algorithminvariant}, the algorithm increases the length of $e$ by an $\exp(\epsilon)$ factor, whenever the congestion is increased in the $t$-th iteration.
If $S$ is empty, then there is no triangle $(u, w, v) \in \PTwo$ with $l^{(t)}_{(u, w, v)} < (1 + \epsilon)\alpha^{(t)}$, 
which implies that the shortest triangle now has a length at least $(1+\epsilon)\alpha^{(t)}$.
In this case, the algorithm increases $\alpha^{(t)}$ by a $(1+\epsilon)$ factor, i.e., $\alpha^{(t+1)} \gets (1+\epsilon)\alpha^{(t)}$.


\paragraph{Computing the Primal Solution with a Small Support Size.} It turns out the length function $l_e^{(t)}$ itself, when divided by $\alpha^{(t)}$ is feasible for $\PRIMAL^{(2)}$ (see~\Cref{lemma:feasibilitydirect}).
Let $z_e^{(t)} := l_e^{(t)} / \alpha^{(t)}$.
The primal objective then becomes $\sum_{e} z_e^{(t)} = \Phi(t) / \alpha^{(t)}$.
To compute the smallest primal objective value,
\Cref{alg:lp} selects an iteration $T$ that minimizes $\sum_{e\in E^+\cup E^-} z^{(T)}_e$.

To ensure the primal solution has a small support size, note that, although we will be able to bound the number of negative edges with non-zero flows, it does not necessarily translate to an upper bound on the number of negative edges with non-zero primal values, as an edge with zero flow has a non-zero primal value of $1/\alpha^{(T)}$. To overcome this issue, we set the primal values of negative edges to be $0$ when there is no flow. However, doing so might violate some constraints.
To compensate this, we may re-adjust the primal values of some positive edges by increasing the primal value of \emph{all} positive edges by $1/(2\alpha^{(T)})$. When $\alpha^{(T)}$ is sufficiently large, we can upper bound the total increase of the primal values. To ensure our $\alpha^{(T)}$ is large enough, we show that a high value of $\Phi(T)$ implies a high value of $\alpha^{(T)}$. Then, when taking $T$ to be the iteration with the minimum $\sum_{e\in E^+\cup E^-} z^{(T)}_e$, we restrict $T \geq T_{\min}$, where $T_{\min}$ is the first iteration the potential $\Phi$ grows to be at least $m^3/\epsilon$. 


\medskip

In the following subsections, we will prove that the primal solution $\{z_e\}$ output from \Cref{alg:lp} is both feasible and $(1 + O(\epsilon))$-approximate.
While we have not yet provided details on how to compute the maximal edge-disjoint eligible open triangles, we will discuss the number of iterations needed at the end of this section.
By combining this with a maximal edge-disjoint eligible open triangles algorithm in \Cref{section:maximal-edge-disjoint-triangles}, we are able to derive the bounds on the running time.

\subsection{Feasibility}
We will begin by showing the feasibility of $\{ z^{(t)}_{uv} = {l^{(t)}_{uv}}/{\alpha^{(t)}} \}$ for any $t$.

\begin{lemma}
\label{lemma:feasibilitydirect}
For any iteration $t$, $\{ z^{(t)}_{uv} = {l^{(t)}_{uv}}/{\alpha^{(t)}} \}$ is feasible for $\PRIMAL^{(2)}$.
\end{lemma}
\begin{proof}
\Cref{obser:algorithminvariant} guarantees that $l^{(t)}_{(u, w, v)} \geq \alpha^{(t)}$ for any $(u, w, v) \in \PTwo$, which ensures that all the constraints of $\PRIMAL^{(2)}$ are satisfied.
\end{proof}

However, when setting up the final $\{ z_e\}$, if $e \in E^{-}$ and $l_e^{(T)} = 1$, the algorithm sets $z_e$ to be $0$ in Line~\ref{ln:trunc-ze}, which reduces the value $z_e$ from what it should be by $1/\alpha^{(T)}$.
This may lead to a violation of a triangle $(u, w, v)$'s primal constraint if $(u, v) \in E^{-}$ and $l_{uv}^{(T)} = 1$.
To address this, we increase $z_e^{(T)}$ by $\epsilon/m$ for all positive edge $e\in E^+$.
We will now show that $\alpha^{(T)} \geq m/(2\epsilon)$,
which implies that the reduction to a negative edge's primal variable is at most $2\epsilon /m$.
Therefore, it suffices to increase the $z_e$ values for all positive edges by $\epsilon / m$, as a triangle in $\PTwo$ contains exactly two positive edges.

\begin{lemma}
\label{lemma:feasibilityminedge}
$\alpha^{(T)} \geq m/(2\epsilon)$.
\end{lemma}
\begin{proof}
Since $\Phi({T_{\min}}) \geq m^3/\epsilon$, by an averaging argument, there exists an edge $e \in \binom{V}{2}$ such that $l^{(T_{\min})}_e \geq \Phi(T_{\min})/\binom{n}{2} \geq m^3/(\epsilon n(n - 1)/2) \geq m/\epsilon$. 
This implies that $\cong^{(T_{\min})}_e > 0$ and at some iteration prior to $T_{\min}$ the algorithm has sent some flow on some triangle $(u,w,v)$ containing $e$. Moreover, using the fact that in each iteration $t$ the algorithm only selects the triangles with length less than $(1+\epsilon) \alpha^{(t)}$ and that $\alpha^{(t)}$ is non-decreasing, we know that the triangles selected at the iteration $T_{\min}-1$ must have length at least $m/(\epsilon\cdot \exp(\epsilon))$.

Hence, whenever $\epsilon < 1/8$ we have
\[
\alpha^{(T)} \ge \frac{m}{\epsilon \cdot \exp(\epsilon) \cdot (1+\epsilon)} \ge  \frac{m}{2\epsilon}. \qedhere
\]
\end{proof}

\begin{lemma}
Algorithm \ref{alg:lp} outputs a feasible solution $\{ z_e \}$ for $\PRIMAL^{(2)}$.
\end{lemma}
\begin{proof}
Let $(u,v,w)$ be an open triangle with $uv,vw \in E^{+}$ and $uw \in E^{-}$. We will show that $z_{uw} + z_{uv} + z_{vw} \geq 1$. First note that if any of $z_{uw}, z^{(T)}_{uv}, z^{(T)}_{vw}$ is greater than 1 then we are done, as at least one of $z_{uw}, z_{uv}$ or $z_{vw}$ will be equal to 1 by Line \ref{ln:finalplus} and \ref{ln:finalminus} of Algorithm \ref{alg:lp}. Otherwise, by Line \ref{ln:finalplus} and \ref{ln:finalminus} of Algorithm \ref{alg:lp}, we have $z_{uw} = z^{(T)}_{uw} + \epsilon / m$, $z_{uv} \geq z^{(T)}_{uv} - 1/\alpha^{(T)}$, and $z_{vw} \geq z^{(T)}_{vw}  - 1/\alpha^{(T)}$. Therefore, 
\begin{align*}
z_{uw} + z_{uv} + z_{vw} &\geq \left(z^{(T)}_{uw} +  \epsilon / m\right) + \left(z^{(T)}_{uv} - 1/\alpha^{(T)}\right) + \left(z^{(T)}_{vw}  - 1/\alpha^{(T)}\right) \\
&\geq z^{(T)}_{uw} + z^{(T)}_{uv} + z^{(T)}_{vw} + \epsilon / m - \epsilon/(2m) - \epsilon/(2m) && \mbox{(by \Cref{lemma:feasibilityminedge})} \\
&\geq z^{(T)}_{uw} + z^{(T)}_{uv} + z^{(T)}_{vw} \geq 1 && \mbox{(by \Cref{lemma:feasibilitydirect})}
\end{align*}
\end{proof}

We now turn our attention to showing that we always maintain a feasible dual $\{y^{(t)}_{(u, w, v)}\}$ throughout the algorithm. First, we show that the potential increases by at most a factor of $\exp(\epsilon)$ in each iteration.
%
\begin{lemma}
\label{lemma:phiincreasealittle}
$\Phi(t)\le \exp(\epsilon) \cdot \Phi(t - 1)$.
\end{lemma}
\begin{proof}
Note that $\Phi(t) = \sum_e l_e^{(t)} \leq  \sum_e \exp(\epsilon) \cdot l^{(t-1)}_e \leq \exp(\epsilon) \cdot \Phi(t-1)$.
\end{proof}


\begin{lemma}
At the beginning of iteration $t$, $\{y^{(t)}_{(u, w, v)}\}$ is a feasible solution to $\DUALTWO$. This holds even for the last iteration $t$ where it does not enter the main body of the loop.
\end{lemma}

\begin{proof}
    It suffices to show that $\cong_e^{(t)} \leq 1$ for any edge $e \in E^{+} \cup E^{-}$ at the beginning of each iteration $t$. Since iteration $t-1$ has been executed, and by the condition of the main loop, we have $\Phi(t-1) \leq {m^{1/\epsilon}}/{\exp(\epsilon)}$.
    By \Cref{lemma:phiincreasealittle}, we have $\Phi(t) \le \exp(\epsilon) \cdot \Phi(t-1) \leq  \exp(\epsilon) \cdot ({m^{1/\epsilon}} / {\exp(\epsilon)}) = m^{1/\epsilon}$ and hence $l_e^{(t)} \leq \Phi(t) \leq m^{1/\epsilon}$. Based on \Cref{obser:algorithminvariant}, $l_e^{(t)} = \big(m^{1/\epsilon}\big)^{\cong_e^{(t)}}$.   Therefore, $\cong_e^{(t)} \leq 1$.
\end{proof}

\subsection{Optimality}

When we compute our primal solution, we first set $z_e^{(T)} = {l^{(T)}_e} / {\alpha^{(T)}}$.
Then, we increase $z_{e}$ for all positive edges by $\epsilon/m$,
which in total impose an extra $\epsilon$ additive quantity to the primal objective.
Let $\OPT$ be the optimal objective value for both $\PRIMAL^{(2)}$ and $\DUALTWO$.
The next lemma gives a bound for $\sum_{e} z^{(T)}_e$.

\begin{lemma} 
\label{lemma:lpapproximatetildex}
Suppose that $2/m\le \epsilon \le 1/10$. Then, $\sum_{e} z^{(T)}_e \leq (1+15\epsilon) \cdot \OPT$. \end{lemma}

\begin{proof}[Proof of \Cref{lemma:lpapproximatetildex}.]
To establish the approximate ratio of the primal solution, it suffices to show that the dual objective $\sum_{p\in\PTwo} y^{(T)}_p$ is within a $(1+O(\epsilon))$-factor 
of the primal objective $\sum_{e} z^{(T)}_e = \Phi(T)/\alpha^{(T)}$.
We first establish the relation between the potential increase $\Phi(t) - \Phi(t - 1)$ and the changes to the dual value within iteration $t-1$:
\begin{align*}
    \Phi(t)-\Phi(t-1)&= \sum_{(u,w,v)\in S} l^{(t)}_{(u,w,v)} (\exp(\epsilon)-1) \\
    &
    \le |S| \cdot (\exp(\epsilon)-1)(1+\epsilon)\alpha^{(t)}\\
    &= \left(\frac{\ln m}{\epsilon^2}\sum_{p\in \PTwo} \Delta y^{(t-1)}_p\right) \cdot (\exp(\epsilon)-1)(1+\epsilon)\alpha^{(t)}\\
    &\leq (1+\epsilon)^2 \frac{\ln m}{\epsilon} \alpha^{(t)} \cdot \sum_{p \in \mathcal{P}_2} \Delta y_p^{(t-1)},\\
\intertext{where $\Delta y_p^{(t-1)} = y_p^{(t)} - y^{(t-1)}_p$ is the flow sent to triangle $p$ at the $(t - 1)$th iteration.
By rearranging the terms, we obtain:}
    \frac{\Phi(t)-\Phi(t-1)}{\alpha^{(t)}} & \le 
    (1+\epsilon)^2 \frac{\ln m}{\epsilon} \cdot \sum_{p \in \mathcal{P}_2} \Delta y_p^{(t-1)}. \tag{1}\label{eq:1}\\
\intertext{In Line~\ref{ln:1-primal-sol} of \Cref{alg:lp}, the algorithm chooses $T$ such that $T \geq T_{min}$ and ${\Phi(T)}/{\alpha^{(T)}}$ is minimized, so} 
    \sum_{e} z_e^{(T)} &= \frac{\Phi(T)}{\alpha^{(T)}} \leq \frac{\Phi(t)}{\alpha^{(t)}} \ \ \ \ \text{for any $t\ge T_{\min}$.}\tag{2}\label{eq:2}\\
\intertext{On the other hand, by \Cref{lemma:phiincreasealittle} and the fact that $\Phi$ is non-decreasing, we have $1 \leq \frac{\Phi(t)}{\Phi(t-1)} \leq \exp(\epsilon)$. When $0 < \epsilon < 1$, we have $\exp(\epsilon) \leq (1 + \epsilon)^2$ and}
    \ln\left(\frac{\Phi(t)}{\Phi(t-1)}\right) &\leq \frac{\Phi(t)}{\Phi(t-1)} - 1  \hspace*{3cm}\mbox{($\ln x\le x-1$ for all $x>0$)}\\ &= \frac{\Phi(t)}{\Phi(t - 1)} \cdot \frac{\Phi(t) - \Phi(t-1)}{\Phi(t)} \\
    &\leq  \exp(\epsilon) \cdot \frac{\Phi(t) - \Phi(t-1)}{\Phi(t)} \\
    &\leq (1 + \epsilon)^2 \frac{\Phi(t) - \Phi(t-1)}{\Phi(t)}.\tag{3}\label{eq:3}
\end{align*}

Therefore, at any iteration $t$, we can bound $\sum_{e} z^{(T)}_e$ by
\begin{align*}
    \sum_{e} z^{(T)}_e \ln\left(\frac{\Phi{(t)}}{\Phi(t-1)}\right) &\leq \frac{\Phi(t)}{\alpha^{(t)}} \cdot (1 + \epsilon)^2 \frac{\Phi(t) - \Phi(t-1)}{\Phi(t)} & \mbox{(by \Cref{eq:3})}\\
    &\leq (1 + \epsilon)^2 \frac{\Phi(t) - \Phi(t-1)}{\alpha^{(t)}} \\
    & \leq (1+\epsilon)^4 \frac{\ln m}{\epsilon} \cdot \sum_{p \in \mathcal{P}_2} \Delta y_p^{(t-1)}. &\mbox{(by \Cref{eq:1})}
\end{align*}
Next, by summing over all $t \geq T_{\min}$, we obtain
\begin{align*}
    \sum_{e} z^{(T)}_e \sum_{t\geq T_{\min}}\ln\left(\frac{\Phi{(t)}}{\Phi(t-1)}\right) \leq (1+\epsilon)^4 \frac{\ln m}{\epsilon} \cdot \left(\sum_{t \geq T_{\min}, p \in \mathcal{P}_2} \Delta y_p^{(t-1)}\right) \tag{4}\label{eq:4}
\end{align*}
It is straightforward to see that the summation of the right-hand side telescopes to at most the current dual objective, which is at most $\OPT$.
To lower bound the left-hand side, we notice that since
$\Phi(T_{\min}-1) < \frac{m^3}{\epsilon}$ and the final $\Phi(T_{last})$ is at least $\frac{m^{1/\epsilon}}{\exp(\epsilon)}$, where $T_{last}$ denotes the last iteration. When $ \frac{2}{m}\leq \epsilon \leq \frac{1}{10}$, we have $\exp(\epsilon)/\epsilon \le m$, thus:
\begin{align*}
\sum_{t\geq T_{\min}}\ln\left(\frac{\Phi{(t)}}{\Phi(t-1)}\right) 
& = \ln\left( \frac{\Phi(T_{last})}{\Phi(T_{\min})} \right) \\
 &\geq \ln \left( \frac{\epsilon \cdot m^{1/\epsilon}}{\exp(\epsilon) \cdot m^3}\right) \\
 &= \left(\frac1\epsilon - 3\right)\ln m - \ln \left(\frac{\exp(\epsilon)}{\epsilon}\right)\\
 &\geq \left({\frac{1}{\epsilon} - 4}\right)\ln m. \tag{5}\label{eq:5}
\end{align*}
Combining all together, when $\frac{2}{m}\leq \epsilon \leq \frac{1}{10}$, we have 
\begin{align*}
   \sum_{e} z^{(T)}_e & =  \frac{\sum_{e} z^{(T)}_e \cdot \big({\frac{1}{\epsilon} - 4}\big)\ln m}{\big({\frac{1}{\epsilon} - 4}\big)\ln m} \\
    &\leq \frac{1}{\big({\frac{1}{\epsilon} - 4}\big)\ln m} \cdot \sum_{e} z^{(T)}_e \cdot \sum_{t\geq T_{\min}}\ln\left(\frac{\Phi{(t)}}{\Phi(t-1)}\right)  & \mbox{(by \Cref{eq:5})}\\
    &\leq \frac{1}{\big({\frac{1}{\epsilon} - 4}\big)\ln m} \cdot (1+\epsilon)^4 \frac{\ln m}{\epsilon} \cdot \left(\sum_{t \geq T_{\min}, p \in \mathcal{P}_2} \Delta y_p^{(t-1)}\right) & \mbox{(by \Cref{eq:4})}\\
    &\leq \frac{(1+\epsilon)^4}{1-4\epsilon} \cdot \OPT \\
    &\le (1+15\epsilon)\cdot \OPT & \mbox{($\epsilon < 1/10$)}
\end{align*}
\end{proof}

Using Lemma \ref{lemma:lpapproximatetildex}, we can show the final output $\{ z_e \}$ is a $(1 + O(\epsilon))$-approximate solution:
\begin{lemma} Suppose that $2/m\le \epsilon \le 1/10$. 
\label{lemma:lpapproximateratio}
    \Cref{alg:lp} outputs $\{z_e \}$ such that $ \sum_{e \in E^{+}\cup E^{-}} z_e \leq (1+16\epsilon) \cdot \OPT$. 
\end{lemma}
\begin{proof}
    By assumption, there must be at least one $(+, +, -)$ triangle, so $\OPT \geq 1$. By \Cref{lemma:lpapproximatetildex}, 
\begin{align*}
    \sum_{e \in E^{+}\cup E^{-}} z_e &\leq \sum_{e} z^{(T)}_e + m \cdot \frac{\epsilon}{m} \\
    &\leq (1 + 15 \epsilon) \OPT + \epsilon \OPT \\
    &\leq (1 + 16\epsilon) \OPT.\qedhere
\end{align*}
\end{proof}

\subsection{Work and Span}
In this section, we will prove the last piece of Lemma \ref{lemma:lpmainlemma}, the parallel running time of Algorithm \ref{alg:lp}.
To begin with, we establish an upper bound on the number of iterations that share the same value of $\alpha$.

\begin{lemma}
\label{lemma:samealphasteps}
For any fixed $\alpha$, there will be at most $R=(3/\epsilon) \ln((1+\epsilon)\alpha)$ iterations such that $\alpha^{(t)} = \alpha$.
\end{lemma}

\begin{proof}
Assume that at some iteration $t$, we have $\alpha(t) = \alpha$. From  \Cref{obser:algorithminvariant}, we know that $l^{(t)}_{(u, w, v)} \geq \alpha$ for any triangle $(u, w, v) \in \PTwo$. 
Our objective is to demonstrate that, after $R=(3/\epsilon) \ln((1+\epsilon)\alpha)$ iterations,
we have $l^{(t+R)}_{(u, w, v)} \geq (1+\epsilon) \alpha$.
Therefore, if $\alpha$ has not been changed, the set of eligible open triangles will be empty and the algorithm has to increase $\alpha^{(t + R)}$.

Assume that $l^{(t)}_{(u, w, v)}\in [\alpha, (1+\epsilon)\alpha)$ for the triangle $(u, w, v)$.
Otherwise, since we never decrease the length function,
we already have $l^{(t+R)}_{(u, w, v)} \geq (1+\epsilon) \alpha$.
If $(u, w, v)$ is ever chosen to an edge-disjoint set $S$ in some iteration $i$, where $i \in [t, t + R)$, then we must have
\begin{align*}
l^{(i+1)}_{(u, w, v)} \ge \exp(\epsilon) \cdot l^{(t)}_{(u, w, v)}\ge (1+\epsilon) \alpha.
\end{align*}

On the other hand, if $(u, w, v)$ has not been chosen into $S$ in any iteration, the algorithm must choose at least one edge on $(u, w, v)$ and increase its length by a factor of $\exp(\epsilon)$ after each iteration.
By the pigeonhole principle, after $R=(3/\epsilon) \ln((1+\epsilon)\alpha)$ iterations, there exists an edge in the triangle $(u, w, v)$ whose length is increased by a factor of $ \exp(\epsilon \cdot R/3)$.
Consequently, the contribution of this edge to $l^{(t+R)}_{(u, w, v)}$ satisfies
\begin{align*}
l^{(t+R)}_{(u, w, v)}
&\ge \exp(\epsilon \cdot R/3) \\
&\geq \exp(\epsilon \cdot \ln((1+\epsilon)\alpha)/\epsilon)\\
&\ge (1+\epsilon)\alpha.
\end{align*}
In either case, we can conclude that $l^{(t+R)}_{(u, w, v)} \geq (1 + \epsilon) \alpha^{(t)}$ for all $(u, w, v) \in \PTwo$.
\end{proof}

To bound the total number of iterations in \Cref{alg:lp}, we need to bound the maximum $\alpha^{(t)}$ and the number of different $\alpha^{(t)}$ values. Since our assumption guarantees that $\OPT(\PRIMAL^{(2)})\geq 1$, we know that there will be at least one triangle in $\PTwo$.
As $\alpha^{(t)}$ is always a lower bound for the shortest triangle at iteration $t$, the maximum possible value of $\alpha^{(t)}$ is $3 \cdot m^{1 / \epsilon}$ when the congestion is $1$. Based on Lemma \ref{lemma:samealphasteps}, there will be $O(\epsilon^{-2}{\log m})$ iterations for any fixed $\alpha$.

Moreover, if $S$ is empty, we increase $\alpha^{(t)}$ by a factor of $(1+\epsilon)$. Therefore, there will be at most $O(\log_{1+\epsilon}(m^{1/\epsilon})) = O(\epsilon^{-2}{\log m})$ different $\alpha^{(t)}$ values.
Combining with \Cref{lemma:samealphasteps}, we obtain the following lemma:

\begin{lemma}
\label{lemma:totalsteps}
In Algorithm \ref{alg:lp}, the total number of iterations is $O(\epsilon^{-4}{\log^2 m})$.\hfill $\square$
\end{lemma}

We can estimate the total number of non-zero terms in the output using Lemma \ref{lemma:totalsteps}.

\begin{lemma}
\label{lemma:numberofnonzeroedges}
The output of \Cref{alg:lp}, $\{z_e\}$, has $O(\epsilon^{-2}{m \log m})$ non-zero values.
\end{lemma}
\begin{proof}
The output $z_e > 0$ if and only if the congestion of the edge $e$ is not zero.
According to Line~\ref{ln:1-sendflow} of the algorithm,
each time the algorithm increases the congestion of a negative edge,
the congestion of \emph{some two} positive edges must be increased by $\epsilon^2/\ln m$ each.
Since there are $m$ positive edges and the congestion is always at most $1$, this implies that at most $\frac12\epsilon^{-2}m\ln m$ negative edges can have non-zero congestion.
\end{proof}

We have not yet specified how to compute the maximal edge-disjoint set $S$. In \Cref{section:maximal-edge-disjoint-triangles}, we prove \Cref{lemma:intro-max-triangle} by giving a parallel algorithm that finds a maximal edge-disjoint set $S$.
\maxtriangle*


We are now ready to prove the main~\Cref{lemma:lpmainlemma} regarding solving $\PRIMAL^{(2)}$. Note that if $\epsilon$ is too small (e.g., $\epsilon \leq 2/m$ in the last section), as we allow $\poly(1/\epsilon)=\poly(m)$ span and work, we can simply run a linear program solver (e.g.~\cite{linearprogramsolver}) to solve (P1) and obtain an approximation ratio of 2.06 by using \cite{chawla2015near}. Hence, we may assume $\epsilon \geq 2/m$. 

\lpmainlemma*
\begin{proof}
By \Cref{lemma:feasibilitydirect} and \Cref{lemma:lpapproximateratio}, \Cref{alg:lp} returns a $(1 + O(\epsilon))$-approximate solution $\{ z_e \}$ for $\PRIMAL^{(2)}$.
To implement \Cref{alg:lp} in the parallel setting, note that by \Cref{lemma:numberofnonzeroedges}, there are at most $O(\epsilon^{-2}{m\log m})$ negative edges of length greater than 1 throughout the algorithm. By \Cref{lemma:intro-max-triangle} it takes $O(m_f^{1.5}\log^3 n)$ work and $O(\log^3 m_f)$ span to compute a maximal edge-disjoint set $S$ of $\PTwo'$, where $m_{f} = O(\epsilon^{-2}{m\log m})$. Therefore, at each iteration, \Cref{alg:lp} takes $O(\epsilon^{-3} m^{1.5}\log^4 n)$ work and $O(\log^3 n)$ span.
By \Cref{lemma:totalsteps}, there are $O(\epsilon^{-4} \log^2 m)$ iterations, so \Cref{alg:lp} takes $O(\epsilon^{-7}{m^{1.5}\log^{6.5} n})$ work and $O(\epsilon^{-4}\log^5 n)$ span in total.
Finally, by \Cref{lemma:numberofnonzeroedges} again, the support size of the returned solution is also $O(\epsilon^{-2} m\log m)$.
\end{proof}



\section{A $2.4$-Approximation Rounding Algorithm}
In this section, first, we present a sequential rounding algorithm that achieves a $2.4$-approximation ratio and then show how to parallelize it. 

Recall that we denote the triangle $(u,w,v)$ by $(uw,wv,vu)$ when we are mapping certain attributes to the edges. Given an assignment $\{ x_e \}_{e \in E^{+} \cup E^{-}}$, when we say a triangle $(uw,wv,vu)$ has edge length $(a, b, c)$, we mean $x_{uw} = a, x_{wv} = b$ and $x_{vu} = c$. 


For an assignment $\{ x_e \}$, we say $\{ x_e \}$ satisfies the \textbf{triangle inequality}, if for all triangles $(uw,wv,vu)$ with edge length $(x_{uw}, x_{vw}, x_{uv})$, we have $x_{uw} + x_{wv} \geq x_{vu}$, $x_{wv} + x_{vu} \geq x_{uw}$, and $x_{vu} + x_{uw} \geq x_{wv}$. For an assignment $\{ x_e \}$, we say $\{ x_e \}$ satisfies the \textbf{\opentrianglequ}, if for all $(+, +, -)$-triangles $(uw,wv,vu)$ with edge length $(x_{uw}, x_{wv}, x_{vu})$, we have $x_{uw} + x_{wv} \geq x_{vu}$.


Our algorithm, shown in Algorithm \ref{alg:seqroundingalgorithm}, takes an assignment $\{ x_e \}_{e\in E^{+} \cup E^{-}}$ satisfying the \opentrianglequ{} as the input. To get an assignment satisfying the \opentrianglequ, we first compute a $(1+\epsilon)$-approximate solution for $\PRIMAL^{(2)}$ and then set $x_{e} = z_{e}$ for all positive edges and $x_{e} = 1 - z_{e}$ for all negative edges. A feasible solution in of $\PRIMAL^{(2)}$ satisfies that for every $uv \in E^{-}$, $uw,wv \in E^{+}$, $z_{vu} + z_{uw} + z_{wv} \geq 1$. This implies $x_{uw} + x_{wv} \geq x_{uv}$, so such $\{x_e\}$ satisfies the \opentrianglequ. Moreover, for all $e$, since $z_{e} \in [0,1]$, we have $x_{e} \in [0,1]$.

 Algorithm \ref{alg:seqroundingalgorithm} is based on the pivot rounding framework of \cite{ACN08, chawla2015near}. The algorithm iteratively selects a random pivot $u$ from the unclustered vertices, forms a cluster by adding each unclustered node $v$ into the cluster with probability $1-p_{uv}$, where $p_{uv}$ is defined as
$$p_{uv} = 
            \begin{cases}
            f^{+}(x_{uv})& \mbox{if } (u,v) \in E^{+},\\
            f^{-}(x_{uv)} & \mbox{if } (u,v) \in E^{-},
            \end{cases}$$
and $f^{+}, f^{-}$ are two functions to be determined. Note that in \Cref{alg:seqroundingalgorithm}, we state the step for choosing a random pivot as choosing the first unclustered node from  a random permutation (Line \ref{ln:randompivot}) for the ease of parallelization in \Cref{sec:parallelround}.

 The main difference between our algorithm and \cite{chawla2015near} is that we choose different $f^{+}, f^{-}$ functions.  This difference arises because in \cite{chawla2015near} the input satisfies the triangle inequality for all types of triangles, while ours only satisfies the \opentrianglequ.

\begin{algorithm}[ht!]
\caption{The sequential rounding algorithm.\\
Input: Graph $G$ and an assignment $\{x_{uv} \}$ satisfying the \opentrianglequ.\\
Output: A partition of $V$.}
\label{alg:seqroundingalgorithm}
\begin{algorithmic}[1]
\Function {\seqrounding}{$G = (V, E), \{ x_{uv} \}$}
    \State Draw a permutation $\pi$ of the vertex set $V$ uniformly at random. 
    \State $V_0 \gets V, t \gets 0$
    \While{$|V_t| > 0$}
    \State \label{ln:randompivot} Let the pivot $w$ be the vertex with the smallest $\pi(w)$ in $V_t$ and set $S_t \gets \{ w \}$ \label{ln:pick_pivot}\\
    \Comment{This step is equivalent to picking the pivot $w \in V_t$ uniformly at random.}
    \State For each vertex $u \in V_t$, add $u$ to $S_t$ with probability $(1 - p_{uw})$ independently.
    \State $V_t \gets V_t \setminus S_t, t \gets t + 1$
    \EndWhile
    \State \Return $\{S_0, S_1, ..., S_{t - 1}\}$
    \EndFunction
\end{algorithmic}
\end{algorithm}

\subsection{Approximation Ratios}
Let $S_w$ be the cluster of $w$ when $w$ is chosen as a pivot. To obtain an approximate ratio of the algorithm, \cite{chawla2015near} consider the following terms for a triangle $(u, v, w)$,

$$cost(u,v \mid w) = 
            \begin{cases}
            Pr[(u \in S_w \textmd{ and } v \not\in S_w) \textmd{ or } (u \not\in S_w  \textmd{ and } v \in S_w )\mid \textmd{$w$ is the pivot}]& \mbox{if } (u,v) \in E^{+},\\
            Pr[u \in S_w \textmd{ and } v \in S_w\mid \textmd{$w$ is the pivot}]  & \mbox{if } (u, v) \in E^{-}
            \end{cases}$$

$$lp(u,v \mid w) = 
\begin{cases}
            x_{uv} \cdot Pr[u \in S_w \textmd{ or } v \in S_w \mid \textmd{$w$ is the pivot}]& \mbox{if } (u,v) \in E^{+},\\
            (1 - x_{uv}) \cdot Pr[u \in S_w \textmd{ or } v \in S_w \mid \textmd{$w$ is the pivot}] & \mbox{if } (u, v) \in E^{-}
            \end{cases}$$

The term $cost(u,v \mid w)$ can be intuitively understood as the cost of the edge $(u,v)$ for Algorithm \ref{alg:seqroundingalgorithm} when $w$ is selected as the pivot. If $w$ is selected as the pivot and $(u,v)$ is a positive edge, then a disagreement occurs if exactly one of $u$ or $v$ is clustered into $S_w$. If $w$ is selected as the pivot and $(u,v)$ is a negative edge, then the disagreement cost is incurred if both $u$ and $v$ are clustered into $S_w$.

On the other hand, the term $lp(u,v \mid w)$ represents the cost of the edge $(u,v)$ for the assignment ${ x_e }$ when $w$ is chosen as the pivot. In a high-level sense, we are trying to charge the actual cost to the objective value of the LP solution, so we will need to make sure that each term in the objective function is charged by at most one triangle throughout the algorithm. Here, we charge the cost contributed by edge $uv$ whenever at least one of $u$ or $v$ is clustered into $S_w$. The contribution of $uv$ to the object value is either $x_{uv}$ or $(1 - x_{uv})$, depending on whether $(u,v) \in E^{+}$ or $(u,v) \in E^{-}$. 

It should be noted that the corresponding probabilities can be expressed by $p_{uw}$ and $p_{vw}$. Once we substitute them into the terms, we obtain the following expressions for $cost(u,v \mid w)$ and $lp(u,v \mid w)$.
$$cost(u,v \mid w) = 
            \begin{cases}
            p_{uw} + p_{vw} - 2p_{uw}p_{vw}& \mbox{if } (u,v) \in E^{+},\\
            (1 - p_{uw})\cdot (1 - p_{vw}) & \mbox{if } (u, v) \in E^{-}
            \end{cases}$$

$$lp(u,v \mid w) =
\begin{cases}
            x_{uv} \cdot (1-p_{uw}p_{vw})& \mbox{if } (u,v) \in E^{+},\\
            (1 - x_{uv}) \cdot (1-p_{uw}p_{vw})& \mbox{if } (u, v) \in E^{-}
            \end{cases}
$$

The analysis considers the cost for a triangle $(u,w,v)$, where each of $u, w$, and $v$ is chosen as a pivot with equal probability, $ALG(uwv)$ and $LP(uwv)$ represents the cost for triangle $(u, w, v)$ for Algorithm \ref{alg:seqroundingalgorithm} and the assignment $\{ x_e \}$, respectively.

$$ALG(uwv) = cost(u, v \mid w) + cost(u, w \mid v) + cost(v,w \mid u)
$$
$$LP(uwv) = lp(u, v \mid w) + lp(u, w \mid v) + lp(v,w \mid u)
$$

\cite{chawla2015near} showed if the ratio between $ALG(uwv)$ and $LP(uwv)$ is upper bounded $\rho$ for every triangle $(u,w,v)$, the output of the Algorithm \ref{alg:seqroundingalgorithm} has an approximation ratio of at most $\rho$ in expectation. More precisely, 

\begin{lemma}[\cite{chawla2015near}]
\label{lemma:triangleimplyapproximateratio}
Fix a set of functions $(f^+, f^-)$ with $f^{+}(0) = f^{-}(0) = 0$. If $ALG(uwv) \leq \rho LP(uwv)$ for every $u, w, v \in V$. Let $ALG$ be the disagreement in clustering Algorithm \ref{alg:seqroundingalgorithm} outputs and $LP = \sum_{e \in E^{+}} x_e + \sum_{e \in E^{-}}(1 - x_e)$, then 
\begin{equation*}
    E[ALG] \leq \rho \cdot LP
\end{equation*}
\end{lemma}


The next question is: What is the best choice of functions $f^{+}$ and $f^{-}$ that minimize $\rho$? \cite{chawla2015near} analyze four different types of triangles (namely, $(+,+,+), (+,+,-), (+,-,-)$, and $(-,-,-)$) and achieve a value of $\rho = 2.06$ by carefully selecting $f^{+}$ and $f^{-}$. Note that for $(+, +, -)$ and $(+, +, +)$ triangles, the triangle inequality is necessary in order to obtain such an approximation ratio with respect to the functions they have designed. As we only have the \opentrianglequ~for $(+, +, -)$ triangles, we will need to come up with different $f^{+}$ and $f^{-}$ functions.

We will first show how to pick the functions to achieve such a 2.4 approximation ratio when the solution satisfies the \opentrianglequ{}. Then, we will show that under the framework of \cite{chawla2015near}, the 2.4 factor is the best ratio we can achieve when the solution does not satisfy the triangle inequality for all the triangles, but only the \opentrianglequ~for $(+,+,-)$ triangles.


\begin{lemma}
Fix $(f^{+}, f^{-})$ as 
$$f^{+}(x) =
\begin{cases}
            1.2 x& \mbox{if } x \leq \frac{5}{6},\\
            1& \mbox{if } x \geq \frac{5}{6}
            \end{cases}
$$
and $f^{-}(x) = x$. For any $\{x_{e}\}$ such that $x_{uw} + x_{wv} \geq x_{uv}$ holds for any $(u, w), (v, w) \in E^{+}$, $(u, v) \in E^{-}$, we have $ALG(uwv) \leq 2.4 \cdot LP(uwv)$.
\end{lemma}
\begin{proof}



To show that our chosen functions $f^{+}(x)$ and $f^{-}(x)$ yield a 2.4 approximation ratio, we will conduct a case-by-case analysis based on different types of triangles. It is worth noting that \cite{chawla2015near} has already established the ratio for $(-,-,-)$ and $(+, -, -)$ triangles even when the solution does not obey the triangle inequality.

\begin{lemma}[\cite{chawla2015near}]
Fix $f^{-}(x) = x$, we have $ALG(uwv) \leq LP(uwv)$ for all $(-,-,-)$ triangles.
\end{lemma}

\begin{lemma}[\cite{chawla2015near}]
Fix $f^{-}(x) = x$, if $f^{+}(x) \leq 2x$ for $x \in [0, 1]$, then we have $ALG(uwv) \leq 2LP(uwv)$, for all $(+,-,-)$ triangles.
\end{lemma}

Let $(a, b, c)$ denote the edge lengths of a triangle $(u, w, v)$, that is, $x_{uw} = a$, $x_{vw} = b$ and $x_{uv} = c$. Define the function $\mathcal{C}(a, b, c)$ as follows: 
$$
    \mathcal{C}(a, b, c) = ALG(uwv) - 2.4LP(uwv)
$$

We begin by showing that for $(+, +, -)$ triangles with edge weights $(a, b, c)$ that satisfies the \opentrianglequ, we have $\mathcal{C}(a, b, c) \leq 0$.

\begin{lemma}
Given our choice of $f^{+}(x)$ and $f^{-}(x)$, for any $(+, +, -)$ triangle with edge weights $(a, b, c)$, where $a + b \geq c$ and $a,b,c \in [0,1]$, we have $ \mathcal{C}(a, b, c) \leq 0$.
\end{lemma}
\begin{proof}
For a $(+, +, -)$ triangle, we have 
\begin{align*}
    \mathcal{C}(a, b, c) = &(1 + 2c - 2cf^{+}(a) - 2cf^{+}(b) + f^{+}(a)f^{+}(b)) - \\
    &2.4(1+a+b-c-bcf^{+}(a)-acf^{+}(b)-(1-c)f^{+}(a)f^{+}(b))
\end{align*}
Depends on whether $a \geq \frac{5}{6}$ or $b \geq \frac{5}{6}$, we have 3 different cases. 
When $a, b\in (\frac{5}{6}, 1]$, we have $f^{+}(x) = 1$ and 
\begin{align*}
    \mathcal{C}(a, b, c) &= (1 + 2c - 2c - 2c + 1) - 2.4(1 + a + b - c - bc - ac - 1 + c) \\
    &= 2 - 2c - 2.4 a - 2.4 b + 2.4ac + 2.4bc \\
    & = (2.4a + 2.4b - 2)c + 2 - 2.4a - 2.4b  && \textmd{ (}2.4 a + 2.4 b - 2 \geq 0 \textmd{ )}\\
    &\leq 2.4a + 2.4b - 2 + 2 - 2.4a - 2.4b \leq 0
\end{align*}

Since $a$ and $b$ are asymmetric, the second case is $a \in [0, \frac{5}{6}]$ and $b \in (\frac{5}{6}, 1]$. We know 
\begin{align*}
    \mathcal{C}(a, b, c) &= (1 + 2c - 2.4ac - 2c + 1.2a) - 2.4(1 + a + b - c - 1.2abc - ac - 1.2a + 1.2ac) \\
    &= -1.4 + 1.68 a - 2.4 b -2.88ac + 2.88abc \\
    &= -1.4 - (-1.68 a + 2.4b) - 2.88ac(1-b) \leq 0
\end{align*}

The last case is when $a, b\in[0, \frac{5}{6}]$, we have $f^{+}(x) = 1.2x$ and
\begin{align*}
    \mathcal{C}(a, b, c) &= (1 + 2c - 2.4ac - 2.4bc + 1.44ab) - 2.4(1 + a + b - c -0.96abc - 1.44ab) \\
    &= -1.4 -2.4 a - 2.4 b + 4.4c + 4.896ab - 2.4ac - 2.4bc + 2.304 abc \\
    &= (4.4 - 2.4a - 2.4b + 2.304ab)c -2.4a - 2.4b + 4.896ab - 1.4
\end{align*}

Since $4.4 - 2.4a - 2.4b + 2.88ab \geq 0$ for $a, b \in [0, \frac{5}{6}]$, $\mathcal{C}(a, b, c)$ will be maximized when $c = min(1, a + b)$. Another point is that $\mathcal{C}(a, b, c)$ is maximized when $a = b$. When $a = b \leq \half$, we have 
\begin{align*}
    \mathcal{C}(a, b, c) &= (4.4 - 2.4a - 2.4b + 2.304ab)c -2.4a - 2.4b + 4.896ab - 1.4 \\
    &= (4.4 - 2.4a - 2.4a + 2.304aa)2a -2.4a - 2.4a + 4.896aa - 1.4 \\
    &= 4.608 a^3 - 4.704 a^2 + 4 a - 1.4 = (a - 0.5) (4.608 a^2 - 2.4a + 2.8) \leq 0
\end{align*}
When $a = b \in [\half, \frac{5}{6}]$, we have 
\begin{align*}
    \mathcal{C}(a, b, c) &= (4.4 - 2.4a - 2.4a + 2.304aa) -2.4a - 2.4a + 4.896aa - 1.4 \\
    &= 7.2 a^2 - 9.6a + 3 = (a - 0.5) (7.2a - 6) \leq 0
\end{align*}
Combining all cases, we have $\mathcal{C}(a, b, c) \leq 0$ for any $(+, +, -)$ triangle whenever $a + b \geq c$.
    
\end{proof}




The remaining case is the $(+, +, +)$ triangles.
\begin{lemma}
Given our choice of $f^{+}(x)$ and $f^{-}(x)$, for any $(+, +, +)$ triangle with edge weights $(a, b, c)$ and $a,b,c \in [0,1]$, we have $\mathcal{C}(a, b, c) \leq 0$.
\end{lemma}

\begin{proof}
Consider a $(+,+,+)$ triangle with edge lengths $(a, b, c)$. We have:
\begin{align*}
    \mathcal{C}(a, b, c) = &2\big(f^{+}(a) + f^{+}(b) +f^{+}(c) - f^{+}(a)f^{+}(b) - f^{+}(b)f^{+}(c) - f^{+}(a)f^{+}(c)\big) - \\
    &2.4\big(a + b + c - cf^{+}(a)f^{+}(b) - af^{+}(b)f^{+}(c) - bf^{+}(a)f^{+}(c) \big)
\end{align*}

Since $\mathcal{C}(a, b, c)$ is symmetric, we can assume without loss of generality that $a \geq b \geq c$. We consider two cases:

{\it Case 1}: At least one of $a, b, c$ is greater than $\frac{5}{6}$. Since $a \geq b \geq c$, $a \geq \frac{5}{6}$. We have:
\begin{align*}
    \mathcal{C}(a, b, c) &= 2\big(1 - f^{+}(b)f^{+}(c)\big) - 
    2.4\big(a + b + c - cf^{+}(b) - af^{+}(b)f^{+}(c) - bf^{+}(c) \big) \\
    & = -(2.4a - 2)(1 - f^{+}(b)f^{+}(c)) - 2.4(b - bf^{+}(c)) - 2.4(c - cf^{+}(b)) \leq 0
\end{align*}

{\it Case 2}: $a, b, c \in[0, \frac{5}{6}]$. In this case, we have:
\begin{align*}
    \mathcal{C}(a, b, c) &= 2\big(1.2a + 1.2b + 1.2c - 1.44ab - 1.44ac - 1.44 bc\big) - 2.4\big(a + b + c - 4.32 abc \big) \\
    &= 2.88(3.6abc - ab - bc - ac) \\
    &= 2.88(ab(1.2c - 1) + bc(1.2a - 1) + ac(1.2b - 1)) \leq 0
\end{align*}
Combining all two cases, we have shown that $\mathcal{C}(a, b, c) \leq 0$ for $(+, +, +)$ triangles.
\end{proof}

Now, we show that the 2.4 approximation ratio is the best we can obtain when the solution does not satisfy the triangle inequality for all the triangles, but only the \opentrianglequ~for $(+,+,-)$ triangles. 
\begin{lemma}
For any $(f^+, f^-)$ with $f^{+}(0) = f^{-}(0) = 0$, there exists a graph $G$ and an assignment $\{x_e\}$ that satisfies the \opentrianglequ~such that there is a triangle $(u,w,v)$ in $G$ with $ALG(uwv) \geq 2.4 LP(uwv)$.
\end{lemma}

To establish the lower bound, let $G$ be a graph containig a $(-,-,-)$ triangle with edge weights $(1,1,1)$, a $(+,+,-)$ triangle with edge lengths $(0.5,0.5,1)$, and a $(+,+,+)$ triangle with edge weights $(0,0,\frac{1}{2})$.

Consider the $(-,-,-)$ triangle with edge weights $(1,1,1)$. Note that $LP(uwv)=0$. If $f^{-}(1)<1$, then $ALG(uwv)>0$, which makes $ALG(uwv)/LP(uwv)$ unbounded. Therefore, we may assume $f^{-}(1)=1$.

Next, we examine the $(+,+,-)$ triangle with edge weights $(0.5,0.5,1)$. Here, we have
$$
    LP(uwv) = 2 \cdot \half \left(1 - f^{+}\left(\half\right) \cdot f^{-}(1)\right) = 1 - f^{+}\left(\half\right)
$$

On the other hand, we have
\begin{align*}
ALG(uwv)&=2\left(f^{+}\left(\frac{1}{2}\right)+f^{-}(1)-2f^{+}\left(\frac{1}{2}\right)\cdot f^{-}(1)\right)+\left(1-f^{+}\left(\frac{1}{2}\right)\right)^2\\
&=\left(1-f^{+}\left(\frac{1}{2}\right)\right)\cdot \left(3-f^{+}\left(\frac{1}{2}\right)\right)
\end{align*}
Hence, we find that $ALG(uwv)=(3-f^{+}\left(\frac{1}{2}\right)) \cdot LP(uwv)$.

Consider now the $(+,+,+)$ triangle with edge weights $(0,0,\frac{1}{2})$. Here, we have $LP(uwv)=\frac{1}{2}$ and $ALG(uwv)=2f^{+}\left(\frac{1}{2}\right)$. Thus, we obtain $ALG(uwv)=4f^{+}\left(\frac{1}{2}\right)LP(uwv)$.

Combining the above two equations, we obtain the inequality
$$
    ALG(uwv) \geq \min\left(3 - f^{+}\left(\half\right), 4f^{+}\left(\half\right)\right) LP(uwv)
$$
When $3-f^{+}\left(\frac{1}{2}\right)=4f^{+}\left(\frac{1}{2}\right)$, we obtain $f^{+}\left(\frac{1}{2}\right)=0.6$ and $ALG(uwv)\geq 2.4LP(uwv)$.
\end{proof}

\subsection{The Parallel Rounding Algorithm}\label{sec:parallelround}
In Algorithm \ref{alg:seqroundingalgorithm}, after a pivot $u$ is chosen, each unclustered node $v$ tries to join the cluster of $u$ with probability $1 - p_{uv}$. To parallelize Algorithm \ref{alg:seqroundingalgorithm}, first, we consider an equivalent sequential algorithm, Algorithm \ref{alg:sequentialroundingalgorithm2}. In this algorithm, instead of revealing the randomness of all the edges incident to $u$ after the pivot $u$ is chosen, we reveal all such randomness at the beginning of the algorithm, before we started to perform any pivoting steps. This can be thought as first constructing an instance $G' = (V, E'^{+}, E'^{-})$ where each edge $uv$ is labelled as $+$ with probability  $1 - p_{uv}$  and labelled as $-$ with probability $p_{uv}$. 
Running the standard \pivot{} algorithm of \cite{ACN08} on $G'$ will produce exactly the same output as if we run Algorithm \ref{alg:seqroundingalgorithm} directly, if we use the same randomness for $p_{uv}$ in both algorithms. In sum, we can pre-round the assignments $\{x_e\}$ into an instance $G'$ and then the remaining step is to perform the \pivot{} algorithm.

\begin{algorithm}[ht!]
\caption{The sequential \pivot{} algorithm with pre-rounding.\\
Input: Graph $G$ and an assignment $\{x_{uv} \}$ satisfying the \opentrianglequ.\\
Output: A partition of $V$.}
\label{alg:sequentialroundingalgorithm2}
\begin{algorithmic}[1]
\Function {\textsc{SeqPreRounding}}{$G = (V, E), \{ x_{uv} \}$}
    \For{$(u, v)$ such that $p_{uv}< 1$}
        \State add $(u, v)$ to $E'^{+}$ with probability $1 - p_{uv}$
    \EndFor
    
    \State Draw a permutation $\pi$ of the vertex set $V$ uniformly at random. 
    \State $V_0 \gets V, t \gets 0$
    \While{$|V_t| > 0$}
    \State Let the pivot $w$ be the vertex with the smallest $\pi(w)$ in $V_t$ and set $S_t \gets \{ w \}$
    \State For $u \in V_t$ such that $(u, w) \in E'^{+}$,  add $u$ to $S_t$
    \State $V_t \gets V_t \setminus S_t, t \gets t + 1$
    \EndWhile
    \State \Return $\{S_0, S_1, ..., S_{t - 1}\}$
    \EndFunction
\end{algorithmic}
\end{algorithm}

To parallelize Algorithm \ref{alg:sequentialroundingalgorithm2}, note that the \pivot{} algorithm of \cite{ACN08} is known to be implementable efficiently in the parallel setting \cite{blelloch2012greedy,FischerN20}. The observation was that we can perform multiple steps of Algorithm \ref{alg:sequentialroundingalgorithm2} in one parallel round as follows. Vertices whose $\pi$-values are local minimum serve as the pivots. All non-pivot nodes then join the neighboring pivot with the smallest $\pi$-value.  For completeness, we give the description of our parallel algorithm in Algorithm \ref{alg:parallelroundingalgorithm}.

\begin{algorithm}[ht!]
\caption{The parallel rounding algorithm.\\
Input: Graph $G$ and an assignment $\{x_{uv} \}$ satisfying the \opentrianglequ.\\
Output: A partition of $V$, $\mathcal{S}$.}
\label{alg:parallelroundingalgorithm}
\begin{algorithmic}[1]
\Function {\textsc{ParallelRounding}}{$G = (V, E), \{ x_{uv} \}$}
    \For{$(u, v)$ such that $p_{uv}< 1$}
        \State add $(u, v)$ to $E'$ with probability $1 - p_{uv}$
    \EndFor
    \State $G'\leftarrow (V, E')$
    \State Draw a permutation $\pi$ of the vertex set $V$ uniformly at random. \label{ln:4-draw}
    \While{$|V'| \geq 1$}
        \State Let $W = \{u \in G' \mid \pi(u) < \pi(v) \mbox{ for every $(u,v) \in G'$} \}$. \\ \label{ln:4-earlier-nbr} \Comment{$W$ is the set of vertices in $G'$ with no earlier neighbors.}
        \State For $w \in W$, set $S(w) \gets \{ w \}$
        \For{$v \in V(G') \setminus W$}
            \State Let $w = \arg\min_{u \in W, (u, v) \in G'} \pi(u)$ \label{ln:4-smallest-nbr}
            \State $S(w) \gets S(w) \cup \{ v \}$
            \\ \Comment{every non-pivot vertex $v$ joins the adjacent pivot with the smallest $\pi(\cdot)$-value, if it exists.}
        \EndFor
        \State $G' \gets G' \setminus \bigcup_{w \in W} S(w)$ 
        \State $\mathcal{S} \gets \mathcal{S} \cup (\bigcup_{w \in W} \{ S(w) \})$
    \EndWhile
    \State \Return $\mathcal{S}$
    \EndFunction
\end{algorithmic}
\end{algorithm}

Note that the pivots chosen throughout the algorithm are exactly the vertices that comprise the greedy maximal independent set (MIS) induced by the permutation $\pi$ in $G'^{+} = (V, E'^{+})$. \cite{FischerN20} showed such a process terminates $O(\log n)$ rounds. We can see that Algorithm \ref{alg:parallelroundingalgorithm} produces exactly the same output as Algorithm \ref{alg:sequentialroundingalgorithm2} if they are coupled with the same random permutation $\pi$ and the same randomness for the probability $\{p_{uv}\}$. 

Finally, it is important to note that for our chosen functions $f^{+}$ and $f^{-}$, if $x_{uv} = 1$ then $p_{uv} = 1$. This implies we can ignore the edge $uv$ as it will never be added to $E'^{+}$. 
Therefore, Algorithm \ref{alg:parallelroundingalgorithm} takes $\tilde{O}(m_f)$ work and $\tilde{O}(1)$ span, where $m_f =  |\{ e \in E^{-} \mid x_e < 1\} \cup E^{+}|$ is the number of positive edges plus the number of negative edges such that $x_e < 1$.

Combining all together, we have the following lemma:
\begin{lemma}[A restatement of Lemma \ref{lem:rounding}]
\label{lemma:parallelrounding}
    Given a graph $G^{+} = (V, E^{+})$ and an assignment $\{ x_{e} \}_{e \in E^{+} \cup E^{-}}$ satisfying the \opentrianglequ. Let $LP = \sum_{e \in E^{+}} x_e + \sum_{e \in E^{-}}(1 - x_e)$ and $m_f = |\{ e \in E^{-} \mid x_e < 1\} \cup E^{+}|$. Algorithm \ref{alg:parallelroundingalgorithm} outputs a clustering that is upper-bounded by  $2.4\cdot LP$ in $\tilde{O}(m_f)$ work and $\tilde{O}(1)$ span.
\end{lemma}

By Lemma \ref{lemma:intro-max-triangle} and Lemma \ref{lemma:parallelrounding}, we prove our main theorem as follows:
\begin{proof}[Proof of Theorem \ref{theorem:maintheorem}]
First, we use Algorithm \ref{alg:lp} to compute a $(1+\epsilon)$-approximate solution $\{z_e\}$ of $\PRIMAL^{(2)}$. We then set $x_e = z_e$ for all positive edges and $x_e = 1 - z_e$ for all negative edges. Let $LP = \sum_{e \in E^{+}} x_e + \sum_{e \in E^{-}} (1 - x_e)$. We have $LP = \sum_{e \in E^{+} \cup E^{-}} z_e \leq (1 + \epsilon) \cdot \OPT(\PRIMAL^{(2)})$. By Lemma \ref{lemma:intro-max-triangle}, this step takes $\tilde{O}(\epsilon^{-7}m)$ work and $\tilde{O}(\epsilon^{-4})$ span.

After running Algorithm \ref{alg:parallelroundingalgorithm}, we obtain a clustering whose cost is at most $2.4 \cdot LP = (2.4 + O(\epsilon)) \cdot \OPT(\PRIMAL^{(2)}) \leq (2.4 + O(\epsilon)) \cdot \OPT(\PRIMAL_I)$, where $\OPT(\PRIMAL_I)$ denotes the optimal value of the correlation clustering problem. Since $m_{f} = \tilde{O}(\epsilon^{-2} m)$, by Lemma \ref{lemma:parallelrounding}, this step takes $\tilde{O}(\epsilon^{-2}m)$ work and $\tilde{O}(1)$ span.

The total running cost is $\tilde{O}(\epsilon^{-7}m)$ work and $\tilde{O}(\epsilon^{-4})$ span, which is dominated by the cost for solving $\PRIMAL^{(2)}$.
\end{proof}

\section{Maximal Edge-Disjoint Eligible Open Triangles}
\label{section:maximal-edge-disjoint-triangles}
Recall that we are given a graph $G = (V, E^{+}\cup E_{>1}^{-})$, where $E^{+}$ is the set of positive edges and $E_{>1}^{-}$ is the set of negative edges with non-zero flow (and thus, with length greater than $1$). We let $m_{f} = |E^{+}\cup E_{>1}^{-}|$ and $m = |E^{+}|$. Note that we will always have $m_{f} = O(m \log m/\epsilon^2)$ as implied by \Cref{lemma:numberofnonzeroedges}. We are also given a length function $l: \binom{V}{2} \to [1, \infty)$, with $l(u,v) = 1$ for each $(u,v) \in \binom{V}{2} \setminus (E^{+}\cup E_{>1}^{-})$. Throughout this section, we say that an open triangle $(u, w, v)$ is \emph{eligible} if  $(w,u), (w,v)\in E^+$, $(u,v) \not\in E^{+}$, and $l(w,u)+l(w,v)+l(u,v) < (1+\epsilon)\alpha$. Our task is to compute a maximal edge-disjoint set $S$ of $\PTwo'$, where $\PTwo' := \{(u,w,v) |\ (u, w), (w, v) \in E^{+}, (u, v) \not\in E^{+}, l(u,w,v) < (1+\epsilon)\alpha\}$ is the set of eligible open triangles.

In this section, we will prove \Cref{lemma:intro-max-triangle} by giving a parallel combinatorial algorithm for finding a \maximalsets. The algorithm uses $O(m_{f}^{1.5}\log^3 n)$ work and $O(\log^3 n)$ span. Additionally, we will show that if Boolean matrix multiplication does not have a truly subcubic time combinatorial algorithm, our algorithm is nearly work-optimal. 

\subsection{Parallel \maximalsets}
We first show a natural sequential algorithm for \maximalsets~and why it is difficult to parallelize it. 

\paragraph{The Conflict Graph Approach.}
One may think of constructing a \emph{conflict graph} $\mathcal{G} = (\mathcal{V}, \mathcal{E})$, where $\mathcal{V} = \PTwo'$ is the set of all eligible $(+,+,-)$ triangles and $e \in \mathcal{E}$ means there is a common edge between two eligible triangles. Then once we run an efficient parallel greedy MIS algorithm~\cite{blelloch2012greedy,FischerN20}, we can get \maximalsets.
However, a very subtle point is that despite there being $O(m^{1.5})$ $(+,+,+)$ triangles on the graph $G$, there could be as many as eligible $\Theta(mn)$ $(+,+,-)$ triangles, which is $\Theta(n^3)$ when the graph is dense. Thus, simulating the greedy MIS algorithm directly takes up to $\tilde{O}(mn)$ total work in the worst case.

\paragraph{A Sequential Greedy MIS Algorithm in $O(m_{f}^{1.5})$ Time.}
It is certainly not necessary to enumerate all $O(mn)$ eligible $(+,+,-)$ triangles at once before invoking the MIS computation on the conflict graph $\mathcal{G}$.
Consider the following sequential algorithm:
for each positive edge $(w, u)\in E^+$, the algorithm considers each incident positive edges $(w, v)$ in the order of non-decreasing length.
Sorting the edges is beneficial because once $l(w, u)+l(w, v)+1\ge (1+\epsilon)\alpha$ we are certain that all eligible triangles involving $(w, u)$ were explored.
Hence, it suffices to consider only the edges $(w, v)$ such that $l(w,u)+l(w, v)+1<(1+\epsilon)\alpha$.

Upon considering an edge $(w, v)$, the algorithm checks if the triangle $(u, w, v)$ is eligible.
If $(u, w, v)$ is an eligible $(+, +, -)$ triangle, the algorithm simply adds this triangle to the MIS and removes both edges $(w, u)$ and $(w, v)$ from $E^+$.
Otherwise, we have found an \emph{unwanted} triangle: 
it could be a $(+, +, +)$ triangle, a $(+,+,-)$ triangle whose negative edge has a large length, or a  $(+,+,-)$ triangle whose negative edge has already been added some triangle in the MIS.

In the end of the algorithm, we are able to deduce that for each positive edge $(w, u)\in E^+$, either it belongs to some eligible triangles in the returned MIS, or all triangles involving the edge $(w, u)$ are now unwanted. To analyze the runtime of the sequential algorithm, it suffices to bound the number of triangles that are once considered throughout the execution.
We notice that the number of once-considered triangles is at most the number of unwanted triangles plus the size of the returned MIS.

To bound the total number of unwanted triangles, we observe that ``the third edge'' of the inspected unwanted triangle is either a positive edge, a negative edge that is already taken into the MIS, or a negative edge that has non-zero congestion. Such negative edges are not many! There can only be at most $O(m_{f})$ of them.
Therefore, with the following folklore result, we are able to bound the number of explored unwanted triangles (we call them \emph{alive} triangles) and thus obtain a $O(m_{f}^{1.5})$ time sequential algorithm.

\begin{lemma}[folklore]\label{lem:count-triangles}
Let $X\subseteq {V\choose 2}$ be an arbitrary set of edges. Then there are at most $|X|^{1.5}$ triangles using only edges in $X$.\hfill $\square$
\end{lemma}

\begin{corollary}\label{cor:unwanted-triangles}
At any moment, the number of triangles that are \emph{explored} but \emph{unwanted} is at most $O(m_{f}^{1.5})$. \hfill $\square$
\end{corollary}

\paragraph{Challenges to Parallelization.}
There are some challenges one has to overcome when parallelizing the above sequential algorithm.
The first idea would be to use separate processors for each positive edge (with a direction) $(w, u)\in E^+$.
Imagine that each positive edge has a list of positive edges $(w, v)$ to be explored. 
We say that an edge with a direction $(w, u)\in E^+$ is \emph{active} if there are still unexplored triangles $(u, w, v)$ for $(w, u)$ where $(w, v)\in E^+$.
To simulate the sequential algorithm, each processor on behalf of a positive edge attempts to explore an eligible triangle. However, there could be some issues when multiple eligible triangles are found at the same time.

For example, if a parallel algorithm discovers $O(m)$ eligible triangles (one for each positive edge) at the same time, there could be multiple eligible triangles sharing the same edge.
In this case, only a few triangles can be added to the MIS and the others become either unwanted (if the negative edge is used up) or destroyed (if a positive edge is used up).
This creates a long dependency chain and we will have no guarantee that this algorithm terminates in polylogarithmic time.
To mitigate this situation, it seems that for each positive edge $(w, u)\in E^+$, the algorithm has to consider more than one triangle at a time and run a parallel MIS on a larger set of eligible triangles.
However, if the algorithm discovers too many eligible triangles at a time, the total work may become too large and exceed $\Omega(m_{f}^{1.5})$.

\subsubsection{The parallel algorithm for \maximalsets}
\paragraph{The Trick of Doubling or Reset.}
Fortunately, we can apply a \emph{doubling trick} to the above approach.
The parallel algorithm now executes in \emph{rounds}, and there is a global ``exploration rate parameter'' $r$, initially set to be $1/2$.
In each round, every positive edge explores a bunch of $r$ new triangles.
Next, the algorithm collects all eligible triangles $\mathcal{C}$ and runs a parallel maximal independent set (MIS) algorithm on the conflict graph $\mathcal{G}[\mathcal{C}]$. The eligible triangles added to the MIS are then removed from the graph.
Finally, depending on how many triangles are still alive (explored but unwanted) on the graph, the algorithm either \emph{doubles} the parameter $r$ or \emph{resets} $r$ to be $1$.

Intuitively, in each round, the algorithm explores a set of eligible $(+, +, -)$ triangles as long as its size is within a constant fraction of currently alive triangles.
Since the total number of alive triangles is at most $O(m_{f}^{1.5})$ by \Cref{cor:unwanted-triangles}, the total work of the algorithm can then be $\tilde{O}(R\cdot m_{f}^{1.5})$, where $R$ is the total number of rounds (we will show in~\Cref{lem:triangle-reset-rounds} that $R=O(\log^2n)$).

The algorithm is summarized in \Cref{alg:find-max-triangle}. To describe our algorithm in greater detail, we introduce the following notations and highlight the main idea.

\paragraph{Active Arcs $A^i$.}
For each positive edge $(w, u)\in E^+$, there are two ways to form an eligible open triangle --- either attaching another positive edge incident to $w$ or incident to $u$.
Moreover, it suffices to consider the eligible triangles where $(w, u)$ is the shorter positive edge.
Our parallel algorithm considers these two types of triangles separately.
In particular, at each round $i$, the algorithm maintains an \emph{active arc set} $A^i$ that contains all \emph{arcs} $(w, u)$ such that there are still some positive edges incident to $w$ not being explored yet from the viewpoint of $(w, u)$.
An arc becomes inactive if all eligible triangles are explored or the associated edge is removed from $E^+$.

\paragraph{Alive Triangles $\mathit{alive}(w, u)$.} As we mentioned before, an \emph{alive} triangle is an explored triangle but unwanted (either ineligible or some edge that already belongs to a triangle in the output set). Given an active arc $(w, u)\in A^i$, we define the set $\mathit{alive}(w, u)$ to be the triangles $(u, w, v)$ that has been explored so far, such that $l(w, u) < l(w, v)$ (or $\mathit{ID}(u) \leq \mathit{ID}(v)$ if $l(w, u) = l(w, v)$) and $(w, v)\in E^+$ \emph{has not been removed yet}.

\paragraph{Sorted Neighbor Lists $N_w^{\mathrm{static}}, N_w^i$.}
For all active arcs leaving $w$, the lists of positive edges to be explored are all incident to $w$. Hence, it would be convenient for the algorithm to sort the neighbors $N(w)$ by its incident edge length $l(w, v)$ in increasing order, and breaking ties by vertex ID.
The sorted list does not change frequently, as there will be many active arcs accessing it via the exploration.
We keep the sorted list as $N_w^{\mathrm{static}}$ and only update the list in each ``reset''.
In particular, in a round with exploration rate $r$, each active edge $(w, u)\in A^i$ explores the neighbors in  $N_w^{\mathrm{static}}[\mathrm{idx}_{\mathrm{static}}{(w,u)}+r, \ldots, \mathrm{idx}_{\mathrm{static}}{(w,u)}+2r-1]$,
where $\mathrm{idx}_{\mathrm{static}}(w, u)$ is the index of $u$ appearing in $N_w^{\mathrm{static}}$.
We remark that (1) it is possible for an active arc to explore a positive edge that has just been removed in the previous round, and (2) after the ``reset'', the same ineligible triangle may be explored (and become alive) again, but it will not cause a problem for us.

Moreover, to support fast calculation to $|\mathit{alive}(w, u)|$ at the beginning of each round, the algorithm maintains the latest \emph{sorted list of neighbors} $N_w^i$ for each round $i$ and each vertex $w$.

\paragraph{Counting Alive Triangles Faster with $\mathit{cur}(w, u)$.}
To compute $|\mathit{alive}(w, u)|$, the algorithm memoizes the quantity $\mathit{cur}(w, u)$ which is defined to be the edge length $l(w, v)$ of the last explored triangle (and the vertex ID $v$ for tie-breaking).
Since the algorithm maintains the updated sorted list $N_w^i$ in each round, it suffices to perform a binary search on $N_w^i$ to obtain the exact count $|\mathit{alive}(w, u)| = (\mathrm{idx}_{\mathit{cur}}(w, u) - \mathrm{idx}_{\mathit{self}}(w, u))$,
where $\mathrm{idx}_{\mathit{cur}}(w, u) $ is the index of the last explored neighbor (who has not been deleted yet) from the active arc $(w, u)$ in the list $N_w^i$, and $\mathrm{idx}_{\mathit{self}}(w, u)$ is the index of the vertex $u$ itself in $N_w^i$.

We remark that there are definitely more efficient parallel data structures (i.e., $O(1)$ span) that maintain the number of alive triangles.
However, the bottleneck to our algorithm is the MIS part which already has an $O(\log m_f)=O(\log n)$ span.
Thus, we choose a simpler $O(\log n)$ span implementation for  ease of understanding.

\begin{algorithm}[htbp]
\caption{Find any maximal edge-disjoint set of eligible $(+,+,-)$ triangles.}\label{alg:find-max-triangle}
\begin{algorithmic}[1]
\Function{MaximalEligibleTriangle}{$G$}
\State $\mathcal{S}\gets\emptyset$; $i\gets 0$; $r\gets 1/2$.
\State For each $w\in V$, obtain two lists of sorted neighbors $N_w^{\mathrm{static}}, N_w^0\gets N(w)$.\label{ln:5-sort-nbr}
\State For each positive edge, add its two directions to the active set $A^0\gets E^+$.
\State For each edge with direction $(w, u)\in E^+$, set $\mathit{cur}(w, u) = (l(w, u), u)$.
\While{$A^i \neq \emptyset$} \label{ln:5-terminating}
\Statex {\small \textit{\underline{$\triangleright$ Part 1: resetting or doubling the exploration rate.}}}
\For{each active arc $(w, u)\in A^i$ \textbf{parallel}}
\State Locate indices $\mathrm{idx}_{\mathit{cur}}(w,u)$ and $\mathrm{idx}_{\mathit{self}}(w,u)$ in $N_w^i$ using binary search.\label{ln:5-bsearch}
\State Obtain the count $|\mathit{alive}(w, u)| = \mathrm{idx}_{\mathit{cur}}(w,u) - \mathrm{idx}_{\mathit{self}}(w,u)$.
\EndFor
\If{$\sum_{(w,u)\in A^i}|\mathit{alive}(w,u)| < \frac{1}{4} |A^i| \cdot r$} \label{ln:5-sumalive} 
    \State \textbf{``Reset'':} $r\gets 1$.
    \State For each active arc $(w, u)\in E^+$ set $\mathit{cur}(w, u)\gets (l(w, u), u)$.
    \State Update the list of sorted neighbors $N_w^{\mathrm{static}}\gets N_w^i$.
\Else
    \State \textbf{``Double'':} $r\gets 2r$.
\EndIf

\Statex {\small \textit{\underline{$\triangleright$ Part 2: collecting eligible triangles.}}}
\State $\mathcal{C}\gets \emptyset$.
\For{each $(w, u)\in A^i$, $v\in N_w^{\mathrm{static}}[\mathrm{idx}_{\mathrm{static}}{(w,u)}+r, \ldots, \mathrm{idx}_{\mathrm{static}}{(w,u)}+2r-1]$} \label{ln:5-check-elig}
\If{$(u,w,v)$ is eligible and $(u, w), (v, w),(u, v)$ are not marked}
\State $\mathcal{C}\gets \mathcal{C}\cup \{(u,w,v)\}$\label{ln:5-collect-elig-tri}
\EndIf
\EndFor
\State Update $\mathit{cur}(w, u)\gets (l(w, v'), v')$ where $v'$ is the last inspected neighbor for $(w, u)\in E^+$.
\Statex {\small \textit{\underline{$\triangleright$ Part 3: obtaining an MIS from conflict graph.}}}
\State Find an MIS $S\subseteq \mathcal{C}$ in the conflict graph $\mathcal{G}[\mathcal{C}]$.\Comment{See~\Cref{lemma:simulating-mis}.}\label{ln:5-mis}
\For{$(u,w,v)\in S$}
\State Remove edges $(u, w)$ and $(v, w)$ from $E^+$ and {\textbf{mark}} all edges $(u, w), (v, w)$, and $(u, v)$.\label{ln:5-remove-triangles}
\EndFor
\State $\mathcal{S}\gets \mathcal{S}\cup S$.

\Statex {\small \textit{\underline{$\triangleright$ Part 4: update the set of neighbors and active edges.}}}
\State Recompute the list of sorted neighbors $N_w^{i+1}\gets N(w)$.\Comment{Some edges were removed!}\label{ln:5-resort}
\State $A^{i+1}\gets\emptyset$. \Comment{Update $A^{i+1}$ from $A^{i}$.}
\For{$(w, u)\in A^i$ and $(w, u)$ is not marked}
\If{$v := N^{\mathrm{static}}_w[\mathrm{idx}_{\mathrm{static}}(w, u) + 2r]$ exists and $l(w, u) + l(w, v) + 1 < (1+\epsilon)\alpha$}
\State $A^{i+1}\gets A^{i+1}\cup \{(w, u)\}$.
\EndIf
\EndFor
\State $i\gets i+1$. \Comment{Increment the round number.}
\EndWhile
\State \Return $\mathcal{S}$
\EndFunction
\end{algorithmic}
\end{algorithm}

\paragraph{Resetting the Exploration Rate.}
If the count of alive triangles becomes too small, say less than $\frac14|A^i|\cdot r$,
the algorithm can not afford to explore $2r$ more triangles and hence ``resets'' the exploration.
Specifically, the algorithm sets $r\gets 1$, updates the sorted neighboring list $N_w^{\mathrm{static}}$, and resets the search progress $\mathit{cur}(w, u)$ for every active arc.
Notice that two things are not being reset --- the active arcs $A^i$ is still decreasing, and the output set $\mathcal{S}$ is still increasing in size.
In particular, we can show that the number of active arcs, compared with the number of active arcs at the last ``reset'' round, must be reduced by a constant fraction
(\Cref{lem:triangle-reset-rounds}), leading to a polylogarithmic span.

\medskip

We summarize our algorithm in~\Cref{alg:find-max-triangle}. Now, we are ready for the analysis.

\subsubsection{Correctness}
\begin{lemma}[Correctness]\label{lem:max-triangle-correctness}
Let $\mathcal{S}$ be the output of \Cref{alg:find-max-triangle}. Then,
$\mathcal{S}$ is an MIS of the conflict graph $\mathcal{G}$.
\end{lemma}
\begin{proof}
It suffices to show that all eligible triangles are explored, which is indeed guaranteed by Line~\ref{ln:5-terminating}, since at the end of the algorithm $A^i=\emptyset$, which implies that all arcs become inactive.
\end{proof}

\subsubsection{Work and Span}

\paragraph{Span Analysis.}

In \Cref{alg:find-max-triangle}, each outermost loop can be identified as a ``double'' round or a ``reset'' round.
It is clear that between any two ``reset'' rounds there can be at most $\log \Delta = O(\log n)$ ``double'' rounds.
Hence, it suffices to show that the total number of ``reset'' rounds is at most $O(\log m)$.
Let the potential $\Phi$ to be number of active arcs (i.e., $|A^i|$) throughout the execution.
The following lemma states that between any two consecutive ``reset'' rounds, the potential $\Phi$ is dropped by at least a constant fraction.

\begin{lemma}\label{lem:triangle-reset-rounds}
Consider a round that is a ``reset'' round. Let $\Phi'$ be the number of active arcs at the beginning of the previous ``reset'' round, and let $\Phi$ be the number of active arcs at the beginning of this round.
Then, $\Phi' - \Phi \ge \frac{1}{16}\Phi$. That is, at least $1/17\approx 5.88\%$ of the active neighbors were removed during the previous round.
\end{lemma}

\begin{proof}

Let $r$ be the exploration rate at the beginning of this ``reset'' round.
If no positive edge was removed and no active arc exhausted their exploration since the last ``reset'', then, the number of alive triangles, by definition, will be exactly $\Phi'(2r-1)$.
Since this is a ``reset'' round and $\Phi'(2r-1)>\frac14\Phi r$, there must be some alive triangles being destroyed since the last ``reset'' round.

There are only two types of events that can cause an alive triangle to disappear:
\begin{itemize}[itemsep=-3pt]
    \item Type 1 Event: an active arc finishes its exploration and became inactive.
    \item Type 2 Event: a positive edge is removed and marked in some round.
\end{itemize}

We will later show that for each of these events, at most $8r-4$ alive triangles are removed. Let $E_1$ (resp. $E_2$) be the number of type 1 (resp. type 2) events that happened  since the last ``reset'' round. Note that this round is a ``reset'' round, so we have $\sum_{(w, u)\in A^i} |\mathit{alive}(w, u)| < \frac14\Phi r$ and
\begin{align*}
E_1+E_2 &\ge \frac{\Phi'(2r-1)-\frac14\Phi r}{8r-4}\\
&\ge \frac{\frac74r-1}{8r-4}\Phi & \mbox{($\Phi'\ge \Phi$)}\\
&\ge \frac{3}{16}\Phi. & \mbox{($r\ge 1$)}
\end{align*}

On the other hand, note that type 1 and type 2 events are also the only two reasons for the potential to decrease. Each type 1 event reduces the potential by 1 since one active arc is removed. According to Line~\ref{ln:5-remove-triangles} of the algorithm, the number of type 2 event is exactly twice as many as the size of the triangles in an MIS $S\subseteq\mathcal{C}$ across all the rounds since the last ``reset''.
Since these triangles are edge-disjoint, and at the time a triangle is added to $\mathcal{C}$ one of the positive edges must be active.
Hence, we conclude that 
\[
\Phi'-\Phi \ge \max\{E_1, \frac12E_2\} \ge \frac{1}{16}\Phi. \tag{Equality holds whenever $E_1=\frac12E_2$.}
\]

We now aim to show that both type 1 and type 2 events contribute to the reduction of potential by at most $8r-4$ alive triangles. Let us first consider a type 1 event, in which an active arc finishes its exploration and is removed from $A^i$. Note that $\mathit{alive}(w, u) \leq 2r -1$ for any $(w, u) \in A^i$, and being inactive does not affect the counting of other active arc's alive triangles. Thus, the number of alive triangles is reduced by at most $2r-1$.

Next, let us consider a type 2 event, where the algorithm removes $(x,y)\in E^+$ and adds it to $S$. We claim that such an edge destroys at most $8r-4$ alive triangles.
First, suppose that both $(x,y)$ and $(y,x)$ are active. In this case, consider the sets $\mathit{alive}(x,y)$ and $\mathit{alive}(y,x)$, each of which contains at most $2r-1$ triangles. Thus, removing $(x,y)$ destroys at most $4r-2$ alive triangles from $\mathit{alive}(x,y)$ and $\mathit{alive}(y,x)$.

Furthermore, the removal of $(x, y)$ affects the alive triangles that use the $(x, y)$ as the longer positive edge, that is there might be some $v'$ such that $(x,y) \in alive(x, v')$ or $(y,x) \in alive(y, v')$.
However, there can be at most $2^0+2^1+\cdots + r=2r-1$ of active arcs $(x, v')$ such that $(x,y) \in \mathit{alive}(x, v')$. 
Specifically, these active arcs must satisfy $v'\in N_x^{\mathrm{static}}[\mathrm{idx}_{\mathrm{static}}(x, y) - (2r-1), \ldots, \mathrm{idx}_{\mathrm{static}}(x, y) - 1]$. By symmetry, there can be at most $2r-1$ active arcs $(y, u')$ that ever consider $(u', y, x)$ as an alive triangle.
Hence, there will be at most $8r-4$ alive triangles being destroyed for each type 2 event.
\end{proof}

\paragraph{Work Analysis.}
Now we focus on the total work for \Cref{alg:find-max-triangle}.
For Part 1, performing a binary search for each active arc incurs $O(m\log n)$ work per round.
Computing the number of alive triangles incurs $O(m)$ work per round.
Updating the list of sorted neighbors incurs $O(m\log m)$ work per round.
For Part 4, computing $N_w^i$ takes $O(m\log m)$ work per round and updating $A^{i+1}$ takes $O(m)$ work per round.
For Part 2 and Part 3,
we first bound the size of collected eligible triangles in each round.

\begin{lemma}\label{lem:max-triangle-work-per-round}
Within each round, there are at most $O(m_{f}^{1.5})$ eligible $(+,+,-)$ triangles in $\mathcal{C}$. Moreover, the total work that computes $\mathcal{C}$ in a round is also $O(m_f^{1.5})$.
\end{lemma}

\begin{proof}
We first observe that in any round, $\sum_{w, u} |\mathit{alive}(w, u)|=O(m_{f}^{1.5})$ according to \Cref{cor:unwanted-triangles}.
If in a round the parameter $r$ is reset, it is clear that $|\mathcal{C}|=O(m)$.
Otherwise, according to Line~\ref{ln:5-sumalive}, we must have $\frac14|A^i|\cdot r \le  
 \sum_{w,u}|\mathit{alive}(w, u)| = {O}(m_{f}^{1.5})$.
Since each active arc explores at most $2r$ new triangles, at most $|A^i|\cdot (2r) = O(m_{f}^{1.5})$ triangles are added to $\mathcal{C}$ and the work for Part 2 is also $O(m_f^{1.5})$.
\end{proof}

\begin{lemma}\label{lemma:simulating-mis}
Given a non-empty collection $\mathcal{C}$ of triangles, there exists a parallel algorithm that returns an MIS on $\mathcal{G}[\mathcal{C}]$ in 
$O(|\mathcal{C}|\log |\mathcal{C}|)$ work and $O(\log |\mathcal{C}|)$ span with high probability.
\end{lemma}
\begin{proof}
    We use the parallel greedy MIS algorithm, which works as follows. the algorithm selects an ordering of the vertices $\pi$, uniformly at random. In each round, all local minima are added to the independent set and removed from the graph, along with their neighbors. Here, local minima refer to all vertices that appear before their neighbors in the ordering. \cite{FischerN20} showed that this algorithm terminates in $O(\log n)$ rounds with high probability where $n$ refers to the number of vertices in the graph. 

    In our case, we cannot explicitly construct $\mathcal{G}[\mathcal{C}]$ since it contains ${O}(m_{f}^{1.5})$ vertices and up to $O(m_{f}^{3})$ edges.
    However, we can still compute the local minima of $\mathcal{C}$ without constructing $\mathcal{G}[\mathcal{C}]$. Our key observation is that any two triangles of $\mathcal{C}$ are neighbors in $\mathcal{G}[\mathcal{C}]$ if and only if they have a common edge in $G$.
    Thus, we compute the local minima of edges of triangles instead of accessing edges in $\mathcal{G}[\mathcal{C}]$. Specifically, let $local(e) = \min_{e\in (u, w,v), (u, w, v) \in \mathcal{C}} \pi(u, w, v)$. Then, a triangle $(u, w, v)$ is a local minimum if and only if $\pi(u, w, v) = local(u, w) = local(w, v) = local(u, v)$. Once we obtain the local minima of triangles in $\mathcal{C}$, we can remove those nodes and their neighbors by removing all triangles whose edges intersect with those local minima of triangles. Note that each triangle in $\mathcal{C}$ contains 3 edges, so the total amount of work is $O(|\mathcal{C}|)$ and the span is $O(\log |\mathcal{C}|)$.
\end{proof}

\begin{proof}[Proof of Lemma \ref{lemma:intro-max-triangle}]
    By  \Cref{lem:triangle-reset-rounds}, \Cref{lem:max-triangle-work-per-round}, and \Cref{lemma:simulating-mis}, we conclude that the algorithm runs in $O(m_{f}^{1.5}\log^3 n)$ work and $O(\log^3 n)$ span.
\end{proof}

\subsection{A Randomized Reduction from Triangle Detection}\label{subsection:hardness}
One may wonder whether there exists a faster sequential algorithm for the \maximalsets{} problem. We show that, unfortunately, any combinatorial algorithm for this problem must take $\Omega(m_{f}^{1.5-\delta})$ time for some $\delta > 0$, unless the Boolean matrix multiplication problem has a truly cubic combinatorial algorithm. The conjecture on the non-existence of such an algorithm has been commonly used to establish lower bounds in fine-grained complexity theory (see \cite{AW14}).  Unless such a conjecture is refuted, our algorithm is optimal up to a $\polylog n$ factor.

Instead of solving the maximal set of triangles, we consider a relaxed version of the problem, the {\it eligible open triangle detection problem}:

\begin{itemize}
    \item Input: An undirected graph $G = (V, E^{+} \cup E^{-}_{>1})$ with a length function $l:\binom{V}{2} \to [1, \infty]$ where $l(u,v) = 1$ for every edge $uv \in \binom{V}{2} \setminus (E^{+} \cup E^{-}_{>1})$, a non-negative real number $L$. Let $n = |V|$ and $m_{f} = |E^{+} \cup E^{-}_{>1}|$.
    
    \item Output: Output ``Yes'', if there is a triangle $(u, w, v)$ such that $(u, w), (w, v) \in E^{+}, (u, v) \not\in E^{+}$, and $l(u,v)+l(v,w)+l(v,u) < L$. Output ``No'', if no such triangle exists.
\end{itemize}

 
\paragraph{Why do we consider the relaxed problem?} Algorithm \ref{alg:lp} requires finding a maximal set of eligible open triangles. If there is an algorithm that finds the maximal set of open eligible triangles in $O(m_{f}^{1.5-\delta})$ time for some constant $\delta > 0$, it can clearly be used to solve the eligible open triangle detection problem also in $O(m_{f}^{1.5-\delta})$ time. 

More generally, the eligible open triangle detection problem captures the key difficulty of our steepest descent algorithm. One may argue that it may not necessarily need to find a maximal set of eligible open triangles but {\it some} open triangles, or solve some even easier tasks. However, note that any algorithm that follows the framework of Algorithm \ref{alg:lp} must send some flow through the (nearly) most beneficial triangle. Thus, it has to find at least one open triangle with some bounded length.




\paragraph{Triangle Detection via Eligible Open Triangle Detection}
Given a graph $\hat{G} = (\hat{V}, \hat{E})$, the (original) triangle detection problem is to determine whether there exists a $(u,v,w) \in \hat{G}$ such that $uv,vw,wu \in \hat{E}$. Williams and Williams~\cite{williams2010subcubic} proved the following hardness result for the triangle detection problem:

\begin{theorem}[\cite{williams2010subcubic}]\label{thm:BMM_triangle}
The following all have truly subcubic (i.e.~$O(n^{3-\delta})$ for some constant $\delta > 0$)  combinatorial algorithms, or none of them do:
\begin{itemize}[itemsep=0pt]
    \item Boolean matrix multiplication (BMM).
    \item Detecting if a graph has a triangle.
\end{itemize}
\end{theorem}

Now we will show a reduction from the triangle detection problem to the eligible open triangle detection problem, with an overhead of $O(n^2)$, as stated by the following lemma:

\begin{lemma}
\label{lem:reductionformaximalset}
Suppose that there is an algorithm $\mathcal{A}$ that solves the eligible open triangle detection problem in $T(n)$ time, there is an algorithm that detects if a graph contains a triangle in $\tilde{O}(T(n) + n^2)$ time, with high probability.
\end{lemma}
\begin{proof}
To detect if a graph $\hat{G} = (\hat{V}, \hat{E})$ has a triangle, we use Algorithm \ref{alg:triangledetection}. The algorithm consists of $O(\log n)$ repetitions. In each repetition, it adds edges $e \in \hat{E}$ to $E^{+}$ with probability $1/2$ and sets the edge weight to 1. If there is a triangle $(u, w, v)$ in $G$, then with probability at least $({3}/{8})$, there will be a triangle such that exactly two of the edges are in $E^{+}$ and the length of the triangle is $3$, since a non-edge in $G$ has length $1$.

For each $uv \in \binom{\hat{V}}{2} \setminus \hat{E}$, we add it to ${E}_{>1}^{-}$ with length $5$. So any triangle containing edges from ${E}_{>1}^{-}$ has length at least $5$. If there is no triangle in $\hat{G}$, then any open triangle in $G$ has a length of at least 5. Algorithm $\mathcal{A}$ sets $L$ to 4 and so it will always output ``No''. 

Since we repeat such a process for $O(\log n)$ times, if there is a triangle in $\hat{G}$, then with high probability, Algorithm \ref{alg:triangledetection} outputs ``Yes''. If $\hat{G}$ contains no triangle, then Algorithm \ref{alg:triangledetection} outputs ``No''. The running time is $\tilde{O}(n^2 + T(n))$ because, in each iteration, it takes $O(n^2)$ time to create the graph and $T(n)$ time to invoke the algorithm $\mathcal{A}$.
\end{proof}

\begin{algorithm}[ht!]
\caption{Triangle detection via eligible open triangle detection.\\
Input: A graph $\hat{G} = (\hat{V}, \hat{E})$.\\
Output: ``Yes'' if there is a triangle $(u, w, v) \in G$. Otherwise, output ``No'' .\\
Oracle: $\mathcal{A}(G,L)$, which returns ``Yes'' if there exists an eligible open triangle of length less than $L$ in $G$, and ``No'' otherwise.} 
\label{alg:triangledetection}
\begin{algorithmic}[1]
\Function {\textsc{TriangleDetection}}{$\hat{G} = (\hat{V}, \hat{E})$}

    \Loop{ $O(\log n)$ times}
    \State $V \gets \hat{V}$, $E^{+} \gets \emptyset,E^{-}_{>1} \gets \emptyset$.
    \State For $uv \in \binom{\hat{V}}{2} \setminus \hat{E}$, add $uv$ to ${E}_{>1}^{-}$, set $l(u,v) = 5$.
    \State For $e \in \hat{E}$, with $1/2$ probability, add $e$ to ${E}^{+}$ and set $l(e) = 1$. 
    \State $r \gets \mathcal{A}(G=(V, {E}^{+} \cup {E}_{>1}^{-}, l), L = 4)$.
    \State If $r$ is ``Yes'', \Return ``Yes''.
    \EndLoop
    \State \Return ``No''.
    \EndFunction
    
\end{algorithmic}
\end{algorithm}

\begin{corollary}
    For any constant $0 <\delta < 1$, no combinatorial algorithm can solve the \maximalsets~problem in $O(m_{f}^{1.5 - \delta})$, unless there is a combinatorial algorithm solving Boolean matrix multiplication in truly subcubic time.
\end{corollary}

\begin{proof}
Suppose to the contrary there is an algorithm that solves the \maximalsets~problem in $O(m_{f}^{1.5 - \delta})$ time. Then by our previous argument, it can be used to solve the eligible open triangle detection problem in $O(m_{f}^{1.5 - \delta})$ time. Since $m_{f} \leq n^2$, this algorithm runs in $O(n^{3 - 2\delta})$ time. By \Cref{lem:reductionformaximalset}, the triangle detection problem can be solved in $\tilde{O}(T(n) + n^2) = \tilde{O}(n^{3 - 2\delta} + n^2)$ time. This in turns implies there is a truly subcubic time algorithm for Boolean matrix multiplication by \Cref{thm:BMM_triangle}. A contradiction occurs.
\end{proof}

Note that we only show that combinatorial algorithms are unlikely to beat $O(m_{f}^{1.5-\delta})$, which excludes, for example, algebraic algorithms.
However, we argue that even if we allow all types of algorithms, it is still unlikely that we can obtain an algorithm that is closer to linear time in $m_{f}$. Note that, the current best algorithm for triangle detection takes $O(\min(n^{\omega}, m^{\frac{2\omega}{\omega + 1}}))$ time~\cite{bjorklund2014listing}, where $\omega \sim 2.37$ is the current best exponent of matrix multiplication. If we have any algorithm that solves the \maximalsets{} problem in $O(m_{f}^{\omega/2 - \delta})=O(m_{f}^{1.185 - \delta})$ time for any constant $\delta >0 $, by \Cref{lem:reductionformaximalset}, it would imply an algorithm for the triangle detection that runs in $O(n^{\omega - 2 \delta})$ time. Such a result would be a major improvement for the triangle detection problem.


\bibliographystyle{alpha}
\bibliography{main}

\appendix

\section{Implementation in MPC Model with $\tilde{O}(\epsilon^{-3}m^{1.5})$ Total Space}\label{sec:mpc}
We note that our algorithms (\Cref{alg:lp}, \Cref{alg:parallelroundingalgorithm}, and \Cref{alg:find-max-triangle}) work well also in the Massively Parallel Computation (MPC) model, with a small overhead in rounds.
Let $\delta \in (0, 1)$ be a constant.
In the MPC model, each machine has $O(n^{\delta})$ local storage, and we are allowing $\tilde{O}(\epsilon^{-3}m^{1.5}/n^\delta)$ machines to run at the same time.
The rest of the section devotes to show that our algorithms runs in the MPC model with only a multiplicative $O(1/\delta)$ factor of overhead in rounds.

\paragraph{\Cref{alg:lp} in MPC.}
There are only three places (line~\ref{ln:1-compute-sum}, \ref{ln:1-max-tri}, and \ref{ln:1-primal-sol}) that need attention.
In Line~\ref{ln:1-compute-sum}, computing the sum of all edge lengths requires knowing the congestion of each edge (and the number of non-zero congestion negative edges).
Since the algorithm explicitly stores the $l^{(t)}_{uv}$ values for all non-zero congested edges (by \Cref{lemma:numberofnonzeroedges} there are at most $O(\epsilon^{-2}m\log m)$ of them), Line~\ref{ln:1-compute-sum} can be simulated in
\begin{align*}
O(\log_{n^{\delta}}\epsilon^{-2}m\log m) = O(1/\delta)
\end{align*}
rounds. Notice that we assume $\epsilon\ge 2/m$ so the overhead does not have additional $\poly(1/\epsilon)$ factors.

In Line~\ref{ln:1-max-tri}, see the analysis to \Cref{alg:find-max-triangle} below. Notice that we do not explicitly compute $\PTwo'$.

In Line~\ref{ln:1-primal-sol}, by \Cref{lemma:totalsteps} we know that there are at most $O(\epsilon^{-4}\log^2m)$ iterations, so all $\Phi(t)$ and $\alpha^{(t)}$ values can be stored in one machine. Thus only $O(1)$ rounds are needed to implement Line~\ref{ln:1-primal-sol}.

In case where $\epsilon$ is super small, book keeping all the histories of $\{y^{(t)}_{(u,w,v)}\}$ and $\{l_e^{(t)}\}$ for all $t$ may be expensive.
To save the total amount of space, we do not need to store all the history --- it suffices to keep the values $\{z_e^{(T)}\}$ and update them whenever $T$ gets updated.

\paragraph{\Cref{alg:parallelroundingalgorithm} in MPC.}
In Line~\ref{ln:4-draw}, drawing a permutation can be simulated by choosing uniformly at random real number in $[0, 1]$ for each vertex, this takes $O(1)$ rounds in MPC.
Furthermore, it takes $O(1/\delta)$ rounds to implement Line~\ref{ln:4-earlier-nbr} and Line~\ref{ln:4-smallest-nbr} since for each vertex the algorithm has to gather the smallest values among its neighbors.
All other lines have a constant overhead when implemented in MPC.

\paragraph{\Cref{alg:find-max-triangle} in MPC.}
Sorting the neighbors (Line~\ref{ln:5-sort-nbr} and~\ref{ln:5-resort}) takes $O(1/\delta)$ rounds in MPC.
Simulating a binary search (Line~\ref{ln:5-bsearch}) takes $O(1/\delta)$ rounds.
Counting alive triangles (Line~\ref{ln:5-sumalive}) takes $O(1/\delta)$ rounds.
Inspect and collect the set $\mathcal{C}$ of eligible triangles  (Line~\ref{ln:5-check-elig}) takes $O(1)$ rounds.
Finally, computing an MIS from the conflict graph $\mathcal{G}[\mathcal{C}]$ in Line~\ref{ln:5-mis} takes 
$O((1/\delta)\log n)$ rounds by simulating~\cite{FischerN20}, which has an additional $O(1/\delta)$ factor comparing to the PRAM implementation.

\end{document}